\documentclass[12pt]{iopart}
\usepackage{iopams}
\usepackage{epsf}
\usepackage{graphicx}
\usepackage{psfrag}
\usepackage[]{hyperref}
\usepackage{color}
\newcommand{\ud}{\mathrm{d}}
\newcommand{\uE}[1]{\mathbb{E}\left[ #1 \right] }

\newcommand{\ue}{\mathrm{e}}
\newcommand{\bm}{\bar{m}}
\newcommand{\mcA}{\mathcal{A}}
\newcommand{\mcB}{\mathcal{B}}
\newcommand{\mcC}{\mathcal{C}}
\begin{document}
\title{Probability distribution of returns in the exponential Ornstein-Uhlenbeck model}
\author{G. Bormetti$^{1,2}$, V. Cazzola$^{1}$, G. Montagna$^{3,2,1}$ and O. Nicrosini$^{2,1}$}
\address{$^{1}$ Istituto Universitario di Studi Superiori, Centro Studi Rischio e Sicurezza, Viale Lungo Ticino Sforza,
  56 27100 Pavia, Italy}
\address{$^{2}$ INFN, Sezione di Pavia, Via A.Bassi, 6 27100 Pavia, Italy}
\address{$^{3}$ Universit\`a degli Studi di Pavia, Dipartimento di Fisica Nucleare e Teorica, Via A.Bassi, 6 27100
Pavia, Italy}
\ead{giacomo.bormetti@pv.infn.it, valentina.cazzola@pv.infn.it, guido.montagna@pv.infn.it, oreste.nicrosini@pv.infn.it}
\begin{abstract}
  We analyze the problem of the analytical characterization 
  of the probability distribution of financial returns in the exponential Ornstein-Uhlenbeck 
  model with stochastic volatility. In this model the prices are driven by a Geometric Brownian motion, 
  whose diffusion coefficient is expressed through an exponential function of an hidden variable $Y$
  governed by a mean-reverting process. We derive closed-form  
  expressions for the probability distribution and its characteristic function in two limit cases. 
  In the first one the fluctuations of $Y$ are larger than the volatility normal level, while the second one
  corresponds to the assumption of a small stationary value for the variance of $Y$.

  Theoretical results are tested numerically by intensive use of Monte Carlo simulations. The effectiveness
  of the analytical predictions is checked via a careful analysis of the parameters involved in the numerical
  implementation of the Euler-Maruyama scheme and is tested on a data set of financial indexes. 
  In particular, we discuss results for the German DAX30 and Dow Jones Euro Stoxx 50, finding a good
  agreement between the empirical data and the theoretical description.
\end{abstract}

\vspace{2pc}
\maketitle

%
%
\section{Introduction and motivation}\label{s:introduction}

Since the pioneering work of Bachelier \cite{Bachelier:1900} about the fair price of derivatives contracts exchanged
on the Paris market in the early '900, diffusion processes have been natural candidates for the modeling of 
the stochastic evolution of financial quantities. Few years later Einstein and, independently, Smoluchowski published 
their famous works about the diffusion of Brownian particles inside a suspension \cite{Einstein:1905,Smoluchowski:1906}. 
Their works were later revisited by Langevin \cite{Langevin:1908}, but from a completely different point of view.
Indeed he took the point of view of the single particles and derived a description in terms of differential equations 
governing the microscopic dynamics under the effects of random collisions. What is more important for our work, 
he also focused on the equation driving the velocity of the diffusing particles and showed the existence of a stationary
regime in which the velocity reaches a constant value. He provided the first example of what we currently call in a modern
language a mean-reverting process described by a stochastic differential equation (SDE). 
Similar arguments but in the language of Fokker-Planck partial differential equation can be found in 
\cite{Uhlenbeck_Ornstein:1930,Kramers:1940}.  
In financial context it is quite common to meet quantities that by construction can
not indefinitely diffuse but reasonably fluctuate around a stationary value.  
For example, the modeling of interest rate dynamics developed during the '90s entirely 
deals with mean-reverting rates \cite{Brigo_Mercurio:2006}.\par 
Empirical studies have also shown that the volatility,
the variable that governs the amplitude of returns fluctuations, is not constant, as postulated by Black 
and Scholes in their seminal work about option pricing \cite{Black_Scholes:1973}. Indeed, once a suitable volatility 
proxy is defined, it has been empirically demonstrated that it fluctuates along time switching between regimes 
with higher and lower activity (bursting effect). Moreover, the assumption of a volatility with a stochastic 
nature represents a quite effective mechanism responsible for the excess of kurtosis observed in the empirical 
probability returns distributions. 
A mean-reverting dynamics is therefore a natural candidate for the modeling of volatility.
Indeed in financial literature this idea has been widely exploited, as the introduction of several stochastic 
volatility (SV) models clearly demonstrates 
\cite{Scott:1987,Hull_White:1987,Stein_Stein:1991,Heston:1993,Ball_Roma:1994,Fouque:2000a}. 
The research in this field has also been strongly enhanced by the pioneering work of Carr and Madan 
\cite{Carr_Madan:1999} and, more recently, Lewis \cite{Lewis:2001}, 
through the introduction of Fast Fourier Transform (FFT) numerical 
techniques. Also the physicists' community has independently analyzed and developed models 
to capture the stochastic nature of volatility, with particular emphasis on the comparison of 
the models predictions with real market data. For a review the reader can see \cite{Bouchaud_Potters:2000},
while specific analysis are addressed in \cite{Bacry_Delour_Muzy:2001,Dragulescu_Yakovenko:2002,
Masoliver_Perello:2002,Mas_Per_An:2004,Borland_etal:2005,Borland_Bouchaud:2005,Masoliver_Perello:2006,
Cisana_etal:2007,Bonanno_etal:2007,Valenti_etal:2007}. Our work follows in particular the guidelines of 
\cite{Dragulescu_Yakovenko:2002} and \cite{Masoliver_Perello:2006}.
Both these articles present an analytical characterization of the probability distribution of financial returns
under the assumption of a Geometric Brownian motion coupled through the diffusion coefficient with
a SDE that describes the evolution of the volatility. Drag\u{u}lescu~-~Yakovenko 
\cite{Dragulescu_Yakovenko:2002} and Masoliver~-~Perell\'o \cite{Masoliver_Perello:2006} 
considered two different models, the former authors presenting an analysis for the Heston one 
\cite{Heston:1993}, while the latter focusing on the Scott model \cite{Scott:1987}.
Main results have been derived under the hypothesis that the stochastic process that 
describes the volatility dynamics has 
reached a stationary regime, a reasonable assumption if the relaxation time is negligible with respect to the 
time horizon along with the process evolves.    
Following Masoliver and Perell\'o we consider the Scott or exponential Ornstein-Uhlenbeck 
model, in particular for its well known ability to capture some
stylized facts observed in real market data such as squared-returns autocorrelation, leverage effects 
and multiple time scale properties \cite{Mas_Per_An:2004,Masoliver_Perello:2006,Cisana_etal:2007}. 
However, we extend their analysis relaxing the request of the complete volatility thermalization.
A good reason for our choice is that in many financial applications the involved time horizon is comparable with
the relaxation time of the process. Moreover, some exotic financial instruments do depend on the entire history of 
the process. The characterization of the probability distribution, even if in approximated form, in the out-of-stationary 
regime can provide some interesting insight in the understanding of the process evolution and be of practical 
interest in the field of quantitative finance.

The structure of the article is the following. In 
Section~\ref{s:model} we review the SV model we will deal with providing the reader with the formulation in terms of 
systems of stochastic differential equations. Section~\ref{s:lambda} and \ref{s:beta} share the same structure. Each
of them consists of two subsections, the first one depicts the framework and describes the analytical results, while
the second puts the theoretical results on numerical tests. Firstly we present a closed-form expression for the 
returns probability distribution in the limit case of log-volatility fluctuations higher than the volatility
normal level. After providing evidences that the approximate solution is effective only in the limit of small 
log-volatility variance, we derive an exact closed-formula for the characteristic function but for a linear 
version of the SV model. Again we check the goodness of the results by means of FFT 
algorithms and we test them against the Monte Carlo (MC) simulation of the linear and 
complete models. Section~\ref{s:realdata} is devoted to the comparison of model predictions with real market
data. We discuss in detail the calibration procedure and we apply it on a set of financial indexes from the 
equity sector. We report more relevant results for the German DAX30 index and the Dow Jones Euro Stoxx 50,
a market capitalization-weighted index of 50 European blue-chip stocks from those countries participating
in the European Monetary Union. 
The final Section~\ref{s:conclusions} draws the relevant 
conclusions and suggests some possible applications of the analytical results in the field of financial option 
pricing. 
Some of the results presented in Section~\ref{s:lambda} are partially shared with a paper 
\cite{Perello_Sircar_Masoliver:2008} appeared on the on-line archive during the completion of this work.

%
%
\section{The Model}\label{s:model}

The process we are investigating is a slightly modified version of the stochastic volatility model
originally introduced by Scott \cite{Scott:1987}. In financial and econophysics literature it is commonly 
referred to as an exponential Ornstein-Uhlenbeck process. However, since we allow for a non-null stationary 
mean value for the random variable driving the log-volatility of the returns, in our case it should be more 
correct to speak of an exponential mean-reverting process. A major difference with Scott model is the existence
of a parameter $\rho$, that takes values in $[-1,1]$ and represents the correlation between the two sources of noise 
in the diffusive process. As we will see later, $\rho$ different from zero is crucial to obtain a non trivial 
behavior for the skewness of the returns distribution.

Previous considerations translate in the following coupled SDEs
\begin{equation}\label{eq:dS}
  \eqalign{
    \ud S(t) = \mu S(t) \ud t + m\ue^{Y(t)} S(t) \, \ud W_1(t)\ , \cr
    S(t_0) = s_0 \ ,
  }
\end{equation}
\begin{equation}\label{eq:dY}
  \eqalign{
    \ud Y(t) = \alpha(\gamma - Y(t)) \ud t + k\rho \ud W_1(t)+ k \sqrt{1-\rho^2}\ud W_2(t) \ ,\cr
    Y(t_0) = y_0 \ ,
  }
\end{equation}
where $\ud W_1$ and $\ud W_2$ are two independent Wiener processes, while 
$s_0$, $\mu$, $m$, $y_0$, $\alpha$, $\gamma$, $k$ and $\rho$ are constant parameters that can be calibrated 
on real data. If we define $\sigma(t)=m\ue^{Y(t)}$ and we set $m>0$ 
in Eq.~\eref{eq:dS} we recognize the same dynamics of a 
Geometric Brownian motion, with a constant drift coefficient $\mu$ while $\sigma(t)$, the so-called volatility,
characterizes the amplitude of the fluctuations of $S$. However, the model does not only allow for a time  
varying $\sigma$, but through the exponential function promotes the volatility itself to a random variable 
driven by the dynamics of $Y$. By acting on the value of $k\geq 0$ we can strongly modify the behavior of $Y$.
The case $k=0$ switches off the stochastic nature of $Y$ and it evolves deterministically towards its stationary
value $\gamma$, with a characteristic time $1/\alpha$ ($\alpha>0$). If $k$ is strictly greater than zero, it 
is well known that $Y$ follows a Gaussian process, with
\begin{equation}\label{eq:Ymean}
  \uE{Y}= (y_0-\gamma)\ue^{-\alpha(t-t_0)}+\gamma \stackrel{t\rightarrow +\infty}{\longrightarrow} \gamma
\end{equation}
\begin{equation}\label{eq:Yvar}
  \uE{Y^2}-\uE{Y}^2 = \frac{k^2}{2\alpha}\left[\ue^{-2\alpha(t-t_0)}+1\right] 
  \stackrel{t\rightarrow +\infty}{\longrightarrow} \beta\doteq\frac{k^2}{2\alpha}
\end{equation}
We have used the notation $\uE{\cdot}$ to indicate the expectation with respect to the probability 
distribution of $Y$, so $\uE{Y}$ and $\uE{Y^2}-\uE{Y}^2$ correspond to the usual mean value and variance, 
respectively.
For later convenience, Eq.~\eref{eq:dS} can be simplified into the following
\begin{equation}\label{eq:dX}
  \ud X(t) = -\frac{1}{2}m^2\ue^{2Y(t)} \ud t + m\ue^{Y(t)} \, \ud W_1(t) \ ,
\end{equation}
with boundary condition $X(t_0)=0$, by applying It\^o's Lemma to the centered logarithmic 
return $X(t)=\ln{S(t)} - \ln{S_0}-\mu (t-t_0)$.

Equations \eref{eq:dY} and \eref{eq:dX} summarize the model we want to focus on. In particular, once 
a time horizon $t>t_0$ has been fixed, we are interested in the characterization of the 
returns transition probability distribution $p_X(X(t)| X_0,Y_0)$. The returns distribution
$p_X(X(t))$ can be readily obtained taking the expectation over the initial distribution of $X_0$ and
$Y_0$.   
In \cite{Masoliver_Perello:2006} the authors derive an approximate expression for $p_X$ under the assumption
of a stationary regime for the $Y$ variable, {\it i.e.} normally distributed with mean $\gamma=0$ and variance
$\beta>0$, $p_{Y_0}\sim\mathcal{N}(\gamma=0,\beta)$. 
The further crucial hypothesis assumed by the authors to obtain their results was $\lambda\doteq k/m \gg 1$.
This fact was supported by empirical evidences based on the analysis of daily Dow Jones Index 
returns. In the following section we generalize Masoliver and Perell\'o's result consistently with the boundary 
condition of Eq.~\eref{eq:dY}, $Y_0\sim \delta(Y_0-y_0)$, that is relaxing the request of initial stationarity, 
and allowing for $\gamma$ different from zero. We do not relax the condition for $\lambda$ in order to
follow the same solving strategy of the forward Fokker-Planck equation. However we will show the effectiveness 
of this assumption at the end of the next section providing the results of the numerical MC tests.

Before deriving our results, it can be worth remembering that the distribution of returns is
surely not the only interesting statistical feature of mean-reverting SV models and literature deals 
with other properties, such as squared-returns autocorrelation, leverage effects 
({\it i.e.} past returns and futures volatilities correlation) and multiple time scale properties 
\cite{Masoliver_Perello:2006,Cisana_etal:2007,Fouque:2000b,Bouchaud:2001,Masoliver_Perello:2004}. 
However our interest is guided 
by a possible applications of a closed-form expression for the returns distribution in the field of financial 
option pricing. Indeed, in a forthcoming work in preparation \cite{Bormetti_Cazzola}, 
we will show how the model under consideration here can emerge in a natural way in a risk-neutral description 
of returns dynamics.           

%
%
\section{Limit case I: $k/m \gg 1$}\label{s:lambda}

\subsection{Approximate closed-form expression for the returns distribution}

Given the dynamics \eref{eq:dY} and \eref{eq:dX}, the associated transition probability density 
function $p(x,y|x_0,y_0)$ satisfies the forward Fokker-Planck equation \cite{Risken:1989,Gardiner:1983} 
\begin{equation}\label{eq:FPforward}
  \frac{\partial p}{\partial t}=
  \frac{m^2}{2}\ue^{2y}\frac{\partial p}{\partial x}+
  \frac{m^2}{2}\ue^{2y}\frac{\partial^2 p}{\partial x^2}-
  \alpha\,\gamma\frac{\partial p}{\partial y}+
  \alpha\frac{\partial (yp)}{\partial y}+
  \frac{1}{2}k^2\frac{\partial^2 p}{\partial y^2}+
  \rho mk\frac{\partial^2 (\ue^{y}p)}{\partial x\partial y}
\end{equation}
with initial condition
\begin{equation}\label{eq:FPinitial}
  p(x,y|x_0=0,y_0)\doteq p(x,y|y_0)=\delta(x)\delta(y-y_0)\ .
\end{equation}
Instead of working with $x$, $y$ and $t$, it is more convenient to introduce the dimensionless variables
$$
\tau\doteq k^2(t-t_0), \qquad u\doteq \lambda x \quad \mathrm{and} \quad v\doteq \lambda y\ .  
$$
With respect to $\tau$, $u$ and $v$ Eq.~\eref{eq:FPforward} becomes
\begin{equation}\label{eq:scaledFPforward}
  \frac{\partial p}{\partial \tau}=
  \frac{\ue^{2v/\lambda}}{2\lambda}\frac{\partial p}{\partial u}+
  \frac{\ue^{2v/\lambda}}{2}\frac{\partial^2 p}{\partial u^2}-
  \frac{\alpha\,\gamma\lambda}{k^2}\frac{\partial p}{\partial v}+
  \frac{\alpha}{k^2}\frac{\partial\left(vp\right)}{\partial v}+
  \frac{\lambda^2}{2}\frac{\partial^2 p}{\partial v^2}+
  \rho\lambda\frac{\partial^2\left(\ue^{v/\lambda}p\right)}{\partial u\partial v} \ ,
\end{equation}
with initial condition
\begin{equation}\label{eq:scaledFPinitial}
  p(u,v|y_0)=\delta(u)\delta(v-\lambda y_0)\ .
\end{equation}
The previous relations can be rewritten in terms of the characteristic function  
\begin{equation}\label{eq:characteristic}
 \varphi(\omega_1,\omega_2,\tau|y_0)=\int\ud u\,\ue^{i\omega_1u}\int\ud v\,\ue^{i\omega_2v}p(u,v|y_0)\ .
\end{equation}
The condition $\tau=0$ becomes
\begin{equation}\label{eq:initialphi}
  \varphi(\omega_1,\omega_2,0|y_0)=
  \int\ud u\,\ue^{i\omega_1 u}\delta(u)\,\int\ud v\,\ue^{i\omega_2 v}\delta(v-\lambda y_0)=
  \ue^{i\omega_2\lambda y_0}\ ,
\end{equation}
while Eq.~\eref{eq:scaledFPforward} reduces to
\begin{eqnarray}\label{eq:phiFPforward}
  \frac{\partial \varphi}{\partial \tau}(\omega_1,\omega_2,\tau|y_0)=&
  \!-\!\frac{i\omega_1}{2\lambda}\varphi(\omega_1,\omega_2\!\!-\!\!\frac{2i}{\lambda},\tau|y_0)
  \!-\!\frac{1}{2}\omega_1^2\varphi(\omega_1,\omega_2\!\!-\!\!\frac{2i}{\lambda},\tau|y_0)\nonumber\\
  &\!+\!\frac{i\omega_2\alpha\,\gamma\lambda}{k^2}\varphi(\omega_1,\omega_2,\tau|y_0)
  \!-\!\frac{\omega_2\alpha}{k^2}\frac{\partial \varphi}{\partial\omega_2}(\omega_1,\omega_2,\tau|y_0)\nonumber\\
  &\!-\!\frac{\lambda^2}{2}\omega_2^2\varphi(\omega_1,\omega_2,\tau|y_0)
  \!-\!\rho\lambda\omega_1\omega_2\varphi(\omega_1,\omega_2\!\!-\!\!\frac{i}{\lambda},\tau|y_0)\ .
\end{eqnarray}
In order to solve Eq.~\eref{eq:phiFPforward}, following \cite{Masoliver_Perello:2006},
we try the ansatz
\begin{equation}\label{eq:ansatz}
  \varphi(\omega_1,\omega_2,\tau|y_0)=
  \ue^{
    -A(\omega_1,y_0,\tau)\omega_2^2-B(\omega_1,y_0,\tau)\omega_2-C(\omega_1,y_0,\tau)+{\rm O}(\omega_2^3)
  }
  \ .
\end{equation}
In \ref{appA} we show how to obtain the ordinary differential equations (ODEs) satisfied 
by $A$, $B$ and $C$ in the limit of small 
$\omega_2$ and up to order $1/\lambda$. Once the approximate expression of $C(\omega_1,y_0,\tau)$ is derived, we can
compute the marginal distribution $p_X(x | y_0)$ noticing that
\begin{eqnarray}\label{eq:pXtransform}
  p_X(x | y_0)&=
  \frac{1}{2\pi}\int_{-\infty}^{+\infty}\ue^{-i\omega_1 x}
  \varphi(\omega_1/\lambda,\omega_2=0,\tau|y_0)\ud\omega_1\nonumber\\
  &\simeq\frac{1}{2\pi}\int_{-\infty}^{+\infty}\ue^{-i\omega_1 x}\ue^{-C(\omega_1/\lambda,y_0,\tau)}\ud\omega_1
  \ .
\end{eqnarray}
It is worth observing that $C$ corresponds to the logarithm of the characteristic function, changed by sign. 
Hence we can easily obtain the cumulant of order $n$ of $p_X$
\begin{equation}\label{eq:cumulants}
  k_n=-(-i)^n\left.\frac{\partial^n C(\omega_1/\lambda,y_0,\tau)}{\partial\omega_1^n}\right|_{\omega_1=0}
  \ .
\end{equation}
This suggests that, instead of computing previous integral directly, we can derive an approximated but closed-form 
solution by exploiting the Edgeworth expansion \cite{Cramer:1951}. We report below the expansion considering correction
to the Gaussian distribution only up to fourth normalized Hermite polynomial \cite{Jondeau:2007}
\begin{eqnarray}\label{eq:pXEdgeworth}
  p_X(x | y_0)\simeq&
  \frac{1}{\sqrt{2\pi k_2}}\ue^{-\frac{(x-k1)^2}{2k_2}}\times\nonumber\\
  &\left[
    1 + 
    \frac{k_3}{\sqrt{6}k_2^{3/2}}H_3\left(\frac{x-k1}{\sqrt{k2}}\right)+
    \frac{k_4}{\sqrt{24}k_2^2}H_4\left(\frac{x-k1}{\sqrt{k2}}\right)
    \right]
  \ .
\end{eqnarray}
The explicit expressions of the first four cumulants are given by
\begin{eqnarray}
  \label{eq:k1}
  k_1 =& -\frac{m^2}{2\alpha}\zeta\ , \\
  \label{eq:k2}
  k_2 =& \frac{m^2}{\alpha}\left[ (1+2\gamma)\zeta + 2(y_0-\gamma)(1-\ue^{-\zeta})\right]\ ,\\
  \label{eq:k3}
  k_3 =& 6\rho\frac{m^3k}{\alpha^2}\left[\zeta(1+\gamma)+(y_0-(1+2\gamma))(1-\ue^{-\zeta})
      -(y_0-\gamma)\zeta\ue^{-\zeta}\right] \ ,
\end{eqnarray}
\begin{eqnarray}
  \label{eq:k4}
  k_4 =& 6\frac{m^4k^2}{\alpha^3}\Bigg[ 2\zeta+(1-\ue^{-2\zeta})-4(1-\ue^{-\zeta})\nonumber \\
    &+4\rho^2\left(\zeta+\zeta\ue^{-\zeta}-2(1-\ue^{-\zeta})\right)\nonumber \\
    &-4\rho^2y_0\left(\zeta\ue^{-\zeta}-(1-\ue^{-\zeta})+\frac{1}{2}\zeta^2\ue^{-\zeta}\right)\nonumber \\
    &+4\rho^2\gamma\left(\zeta+2\zeta\ue^{-\zeta}-3(1-\ue^{-\zeta})+\frac{1}{2}\zeta^2\ue^{-\zeta}\right)\Bigg]
  \ ,
\end{eqnarray}
where we have introduced the ancillary variable $\zeta\doteq \alpha(t-t_0)$. 
We have checked the consistency of our results when $t$ approaches $t_0$ with the initial time condition 
\eref{eq:FPinitial}. Indeed, at lowest order in $\zeta$, $k_1$ and $k_2$ scale linearly, 
 while being $k_3=O(\zeta^2)$ and $k_4=O(\zeta^3)$ the skewness $\varsigma\doteq k_3/k_2^{3/2}$ 
and kurtosis $\kappa\doteq k_4/k_2^2$ do not diverge, as expected.

The analytical expression for $C$ given by Eq.~\eref{eq:C} (and from it, by simple derivation, 
the explicit characterization of the cumulants) represents one of the main results of this work.
However, Eq.~\eref{eq:pXEdgeworth} has to be considered with care for various reasons. 
First, it is well known that for an arbitrary choice of $k_1$, $k_2$, $k_3$ and $k_4$ the 
positive definiteness of the truncated Edgeworth expansion is not guaranteed. 
Secondly, when dealing with the problem of probability distribution characterization for the sum of
$n$ random variables ({\it e.g.} identically distributed and with finite moment of every order) higher order terms of the 
Edgeworth expansion scales with the power of $n^{-r/2}$ for suitable $r$. 
For example the term proportional to $H_3$ scales with $n^{-1/2}$, while the terms proportional to $H_4$ and $H_6$ 
(not considered here) scale with $n^{-1}$. But for the case under consideration we have $n=1$, so in general we can
not {\it a priori} estimate the goodness of the approximation. Moreover, the entire procedure previously described 
for the derivation of Eq.~\eref{eq:pXEdgeworth} is strongly based on some approximations.

It is worth noting that similar approximations have been adopted in \cite{Perello_Sircar_Masoliver:2008}, 
where the complete characteristic function is given by Eqs.~(35) and (36) yielding, in their notation, 
the following formula
\begin{eqnarray}
  &&\varphi(\omega/\lambda,\bar{\alpha}t)
  =\exp\Biggl\{-i\omega\mu(t)-\left[\bar{m}^2t+2\vartheta(t,z_0)\right]\frac{\omega^2}{2}
  +i\rho\varsigma(t,z_0)\omega^3 \nonumber \\
  &&\qquad\qquad\qquad\qquad
  +\left(\kappa(t)+\vartheta(t,z_0)^2/2\right)\omega^4+\mbox{O}(1/\lambda^5)\Biggl\}\,,
\end{eqnarray}
where the various arguments entering the exponential are defined in \cite{Perello_Sircar_Masoliver:2008}. 
Following the procedure of retaining the $\bar{m}^2$ term inside the exponential function, 
while performing a Taylor expansion starting from the $\vartheta$ dependent contribution, 
the authors of \cite{Perello_Sircar_Masoliver:2008} get the approximate expression (Eq.~(41) in the same paper)
\begin{eqnarray}
  &&\varphi(\omega/\lambda,\bar{\alpha}t)
  =\exp\Biggl\{-i\omega\mu(t)-\bar{m}^2t\frac{\omega^2}{2}\Biggl\}\left[1-\vartheta(t,z_0)\omega^2
    +i\rho\varsigma(t,z_0)\omega^3\right.\nonumber\\
    &&\left.\qquad\qquad\qquad\qquad
    +\left(\kappa(t)+\vartheta(t,z_0)^2/2\right)\omega^4+\mbox{O}(1/\lambda^5)\right].
\end{eqnarray}
This leads to a distribution of returns containing a $H_2$ Hermite polynomial contribution 
as given by Eq.~(42) in \cite{Perello_Sircar_Masoliver:2008}.

On the other hand, if one follows a different approach expanding the exponential 
starting from the $\omega^3$ term, the approximate characteristic function reads, 
according to the notation of \cite{Perello_Sircar_Masoliver:2008}, as follows
\begin{eqnarray}
  &&\varphi(\omega/\lambda,\bar{\alpha}t)
  =\exp\Biggl\{-i\omega\mu(t)-\left[\bar{m}^2t+2\vartheta(t,z_0)\right]
  \frac{\omega^2}{2}\Biggl\}\left[1+i\rho\varsigma(t,z_0)\omega^3\right.\nonumber\\
    &&\left.\qquad\qquad\qquad\qquad
    +\left(\kappa(t)+\vartheta(t,z_0)^2/2\right)\omega^4+\mbox{O}(1/\lambda^5)\right].
\end{eqnarray}
This leads to the probability distribution of the form of Eq.~\eref{eq:pXEdgeworth}, 
where a $H_2$ polynomial term is absent by construction and incidentally this expansion 
corresponds to Edgeworth series truncated to the fourth Hermite polynomial \cite{Cramer:1951,Jondeau:2007}.

In the next section we will discuss numerical tests based on intensive MC simulations 
in order to check the robustness of our results.

\subsection{Numerical Results}\label{s:FNR}

Standard numerical techniques to simulate random paths from the dynamics Eq.~\eref{eq:dY} and Eq.~\eref{eq:dX}
are widely discussed in literature and a classical reference is given by \cite{Glasserman:2003}. In particular 
Chapter 6 is dedicated to discrete schemes of SDEs suitable to generate paths using Monte Carlo methods. 
In our analysis we implement the Euler-Maruyama scheme. The two crucial parameters to be chosen correspond to the 
discrete time step $\Delta t\doteq (t-t_0)/\mathrm{NSTEP}$ and the total number of MC paths to be generated,
$\mathrm{MCPATHS}$. The time variables are measured in yearly
units, so $t-t_0=1$ corresponds to a diffusion process that evolves for one year. We started with $\Delta t=10^{-1}$
and then decreased it until $\Delta t=10^{-4}$. At the same time we increased $\mathrm{MCPATHS}$ from $10^{3}$ to 
$5\cdot 10^{7}$. We finally fixed $\Delta t=10^{-3}$ and $\mathrm{MCPATHS}=5\cdot 10^{6}$, because we did not 
find a significant improvement with smaller value of $\Delta t$.

All the numerical MC results have been obtained with a private Micro Beowulf Cluster made of 4 nodes, 
each of which is an AMD Athlon 64 Dual Core with 2.00 GHz CPUs and 2.00 GByte of RAM, developed by one of the 
authors (\textit{G.B.}) \cite{Adams_Brom_Layton:2007}. The resort to 64 bit architecture can be useful
to avoid some problem that can arise with high Monte Carlo statistics.

\begin{figure}[ht!]
  \caption{\label{fig:FNRBetaRho-0.9} Returns distributions from MC simulation of the Euler-Maruyama 
    scheme and analytical probability distributions given by Eq.~\eref{eq:pXEdgeworth}. Parameters values as 
    discussed in the text, curves from bottom to top correspond to increasing values of $\beta$: 0.5\%,
    1\%, 2\%, 5\%, 10\%, 25\% and 50\%. 
    Curves have been shifted upwards for the sake of readability.
    Top panel: $\rho=-0.9$ and $t-t_0=0.1$; bottom panel: left tails in semi-logarithmic scale.
  }
  \begin{center}
    \includegraphics[scale = 0.85]{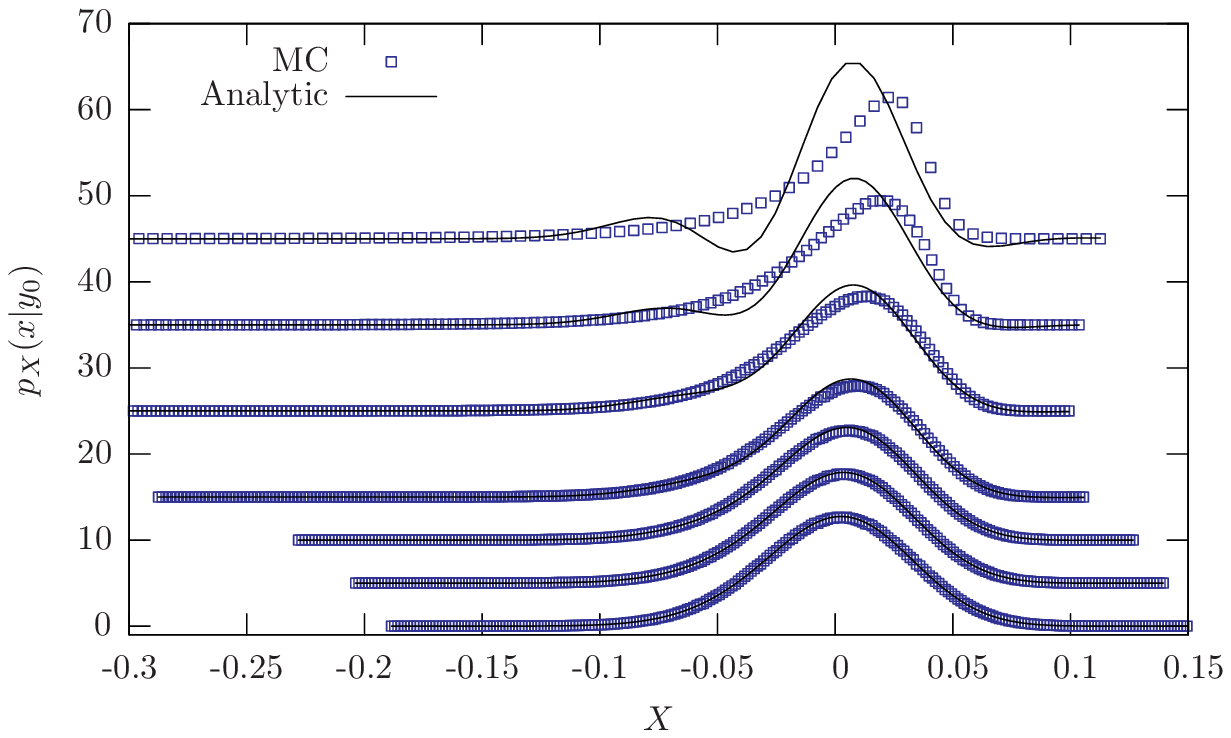}
  \end{center}
  \begin{center}
    \includegraphics[scale = 0.89]{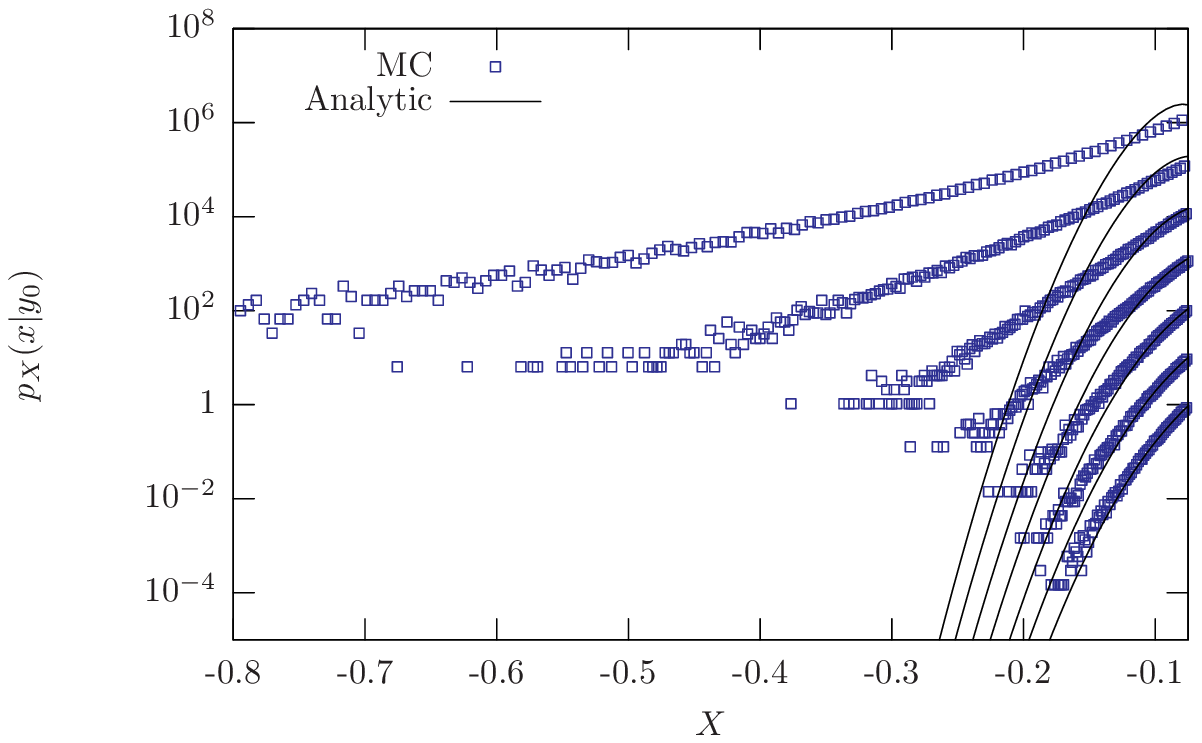}
  \end{center}
\end{figure}

In Fig.~\ref{fig:FNRBetaRho-0.9} we report the returns distributions obtained via MC simulation and the 
analytical predictions given by the Edgeworth expansion~\ref{eq:pXEdgeworth}. 
Curves from bottom to top correspond to increasing values
of $\beta$: 0.5\%, 1\%, 2\%, 5\%, 10\%, 25\% and 50\%, $\rho=-0.9$, $m=0.1$, $\alpha=10$, $\gamma=0$,
$y_0=0$. 
From the relation $\lambda^2=2\cdot 10^3 \beta$, we derive $\lambda=3.16, 4.47, 6.32, 10, 14.14, 22.36, 31.62$, 
that are values consistent with the assumed approximation $\lambda \gg 1$. 
The choice $\gamma=0$ guarantees to preserve the interpretation of $m$ as the normal level of the volatility in 
stationary regime. Indeed, $\gamma\neq 0$ introduces an exponential correction which 
we could easily deal with by a suitable redefinition of $m$ and a linear shift of $Y$. 
The value $m=0.1$ implies a yearly volatility of the $X$ process
of order 10\%, while $\alpha$ equal to 10 corresponds to a relaxation time of 0.1~years.  
In Fig.~\ref{fig:FNRBetaRho-0.9} we consider the time interval $t-t_0=0.1$ in order to test the 
goodness of the analytical approximation in a non-stationary regime.
The agreement is quite good for low values of $\beta$ (and therefore low $\lambda$, but always greater 
than one) and rapidly worsen for higher values for non-stationary regimes. 
The peak of the MC distribution moves rightward and the left tail becomes fatter with respect to the 
analytical prediction as can be clearly seen in the bottom panel of Fig.~\ref{fig:FNRBetaRho-0.9}. 
Indeed for the highest $\beta$ values the Edgeworth expansion starts to oscillate on the tails and becomes negative.  
\begin{figure}[ht!]
  \caption{\label{fig:FNRBetaRho0.5} Details as in the caption of Fig.~\ref{fig:FNRBetaRho-0.9}.
    Top panel: $\rho=0.5$ and $t-t_0=1$; bottom panel: right tails in semi-logarithmic scale.
  }
  \begin{center}
    \includegraphics[scale = 0.85]{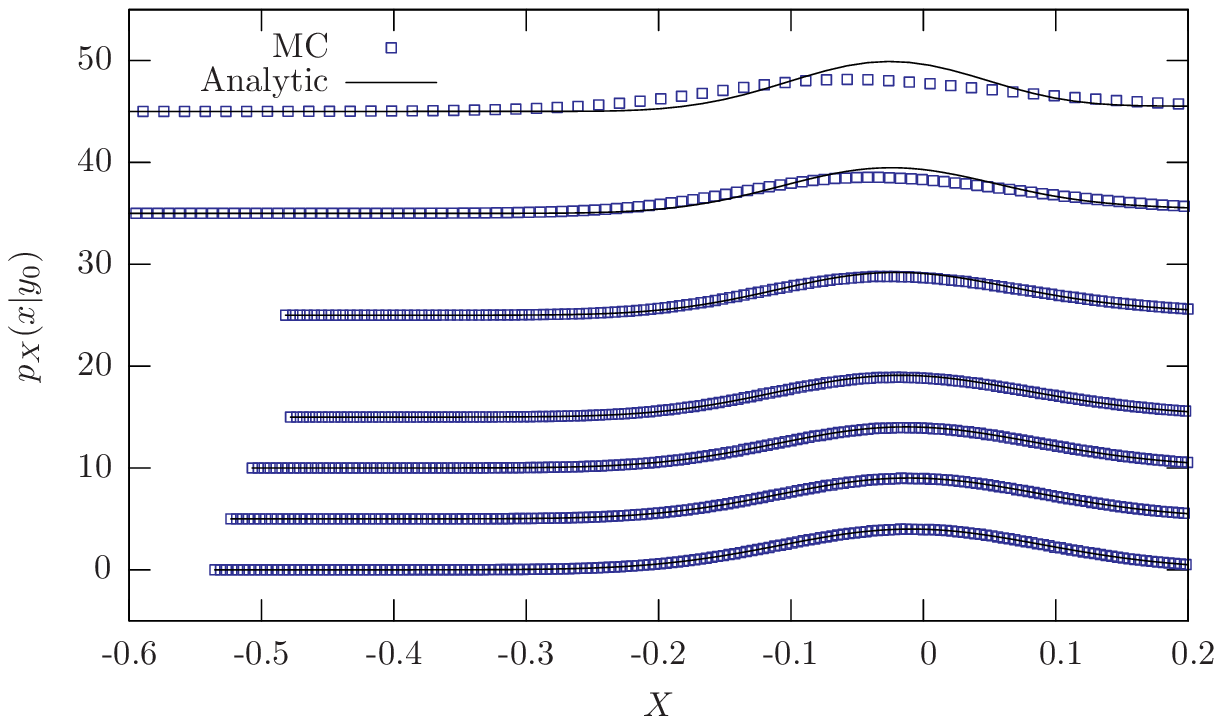}
  \end{center}
  \begin{center}
    \includegraphics[scale = 0.89]{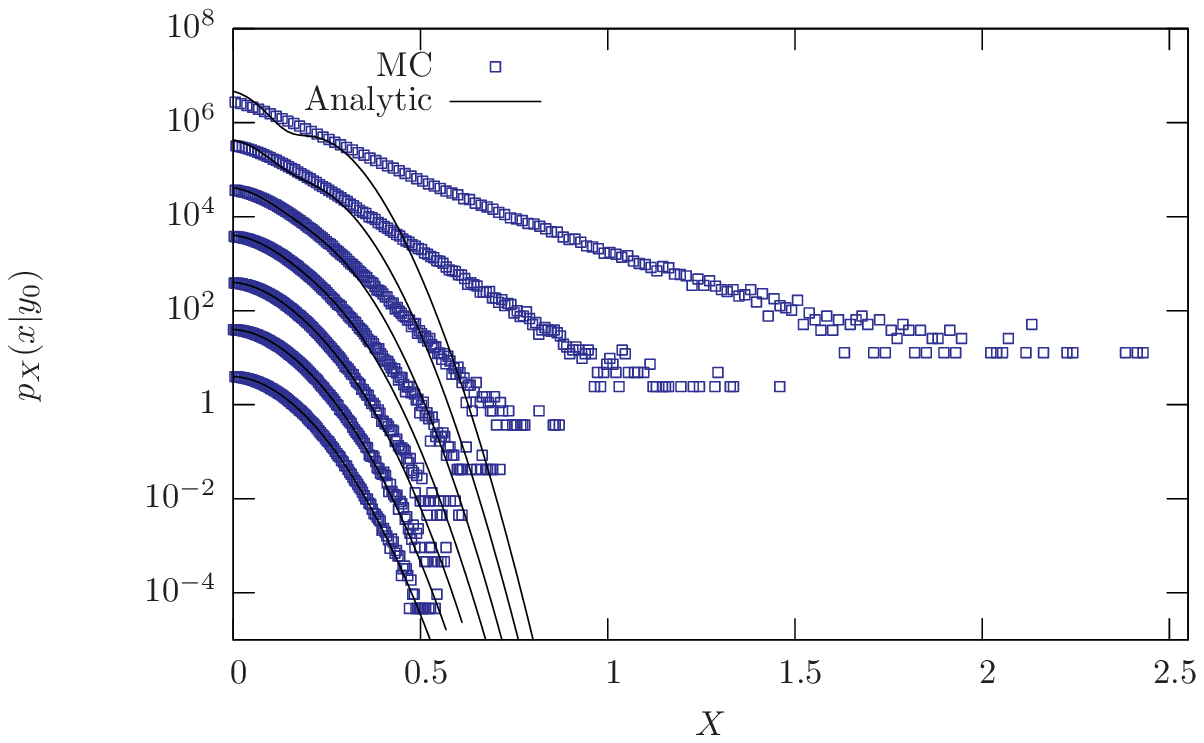}
  \end{center}
\end{figure}
In Fig.~\ref{fig:FNRBetaRho0.5} we show the behavior of the distribution for a positive value of the correlation
coefficient, $\rho=0.5$, and in stationary regime $t-t_0=1$.  After ten relaxation 
times the $Y$ process has completely thermalized 
and again we checked the agreement between MC based results and the Edgeworth expansion.  
As correctly predicted by Eq.~\eref{eq:k3}, $\rho$ governs the sign of the skewness
and the figures confirm that the asymmetry of the distributions is opposite with respect to the previous case.
Again the best agreement corresponds to the lowest $\beta$, while highest values for $\lambda$ 
are not sufficient to improve it. As in Fig.~\ref{fig:FNRBetaRho-0.9}, 
significant differences are present in the distribution tails.
\begin{figure}[ht!]
  \caption{\label{fig:FNRScalingRho-0.9} From top left clockwise: scaling with
    time of mean, variance, kurtosis and skewness. 
    Comparison between numerically estimated values (with MC 95\% confidence level) and theoretical prediction
    (Eq.~\eref{eq:k1}-Eq.~\eref{eq:k4}) for normalized cumulants. $\rho=-0.9$, $\beta=0.5\%, 5\%$ and other parameters 
    as in the text.
  }
  \begin{minipage}[b]{0.5\textwidth} 
    \begin{center}
      \includegraphics[scale = 0.68]{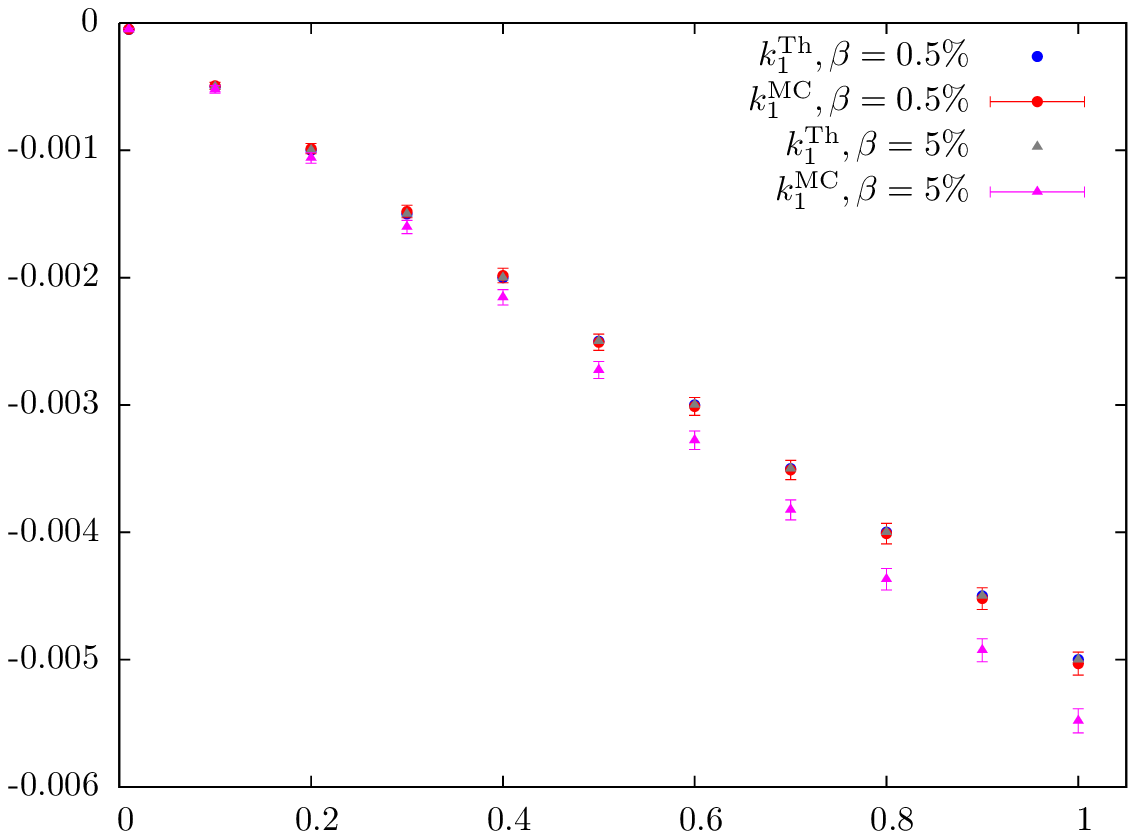}
    \end{center}
  \end{minipage}
  \begin{minipage}[b]{0.5\textwidth}
    \begin{center}
      \includegraphics[scale = 0.68]{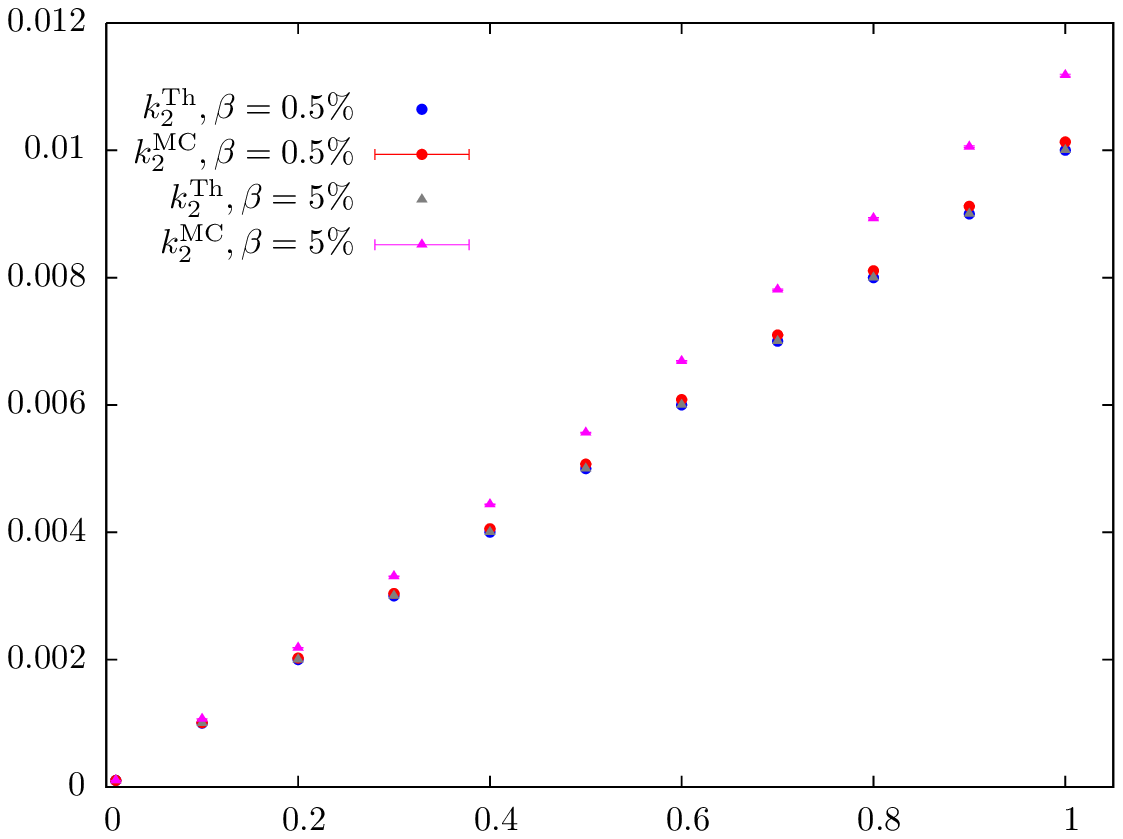}
    \end{center}
  \end{minipage}
  
  \begin{minipage}[b]{0.5\textwidth}
    \begin{center}
      \includegraphics[scale = 0.68]{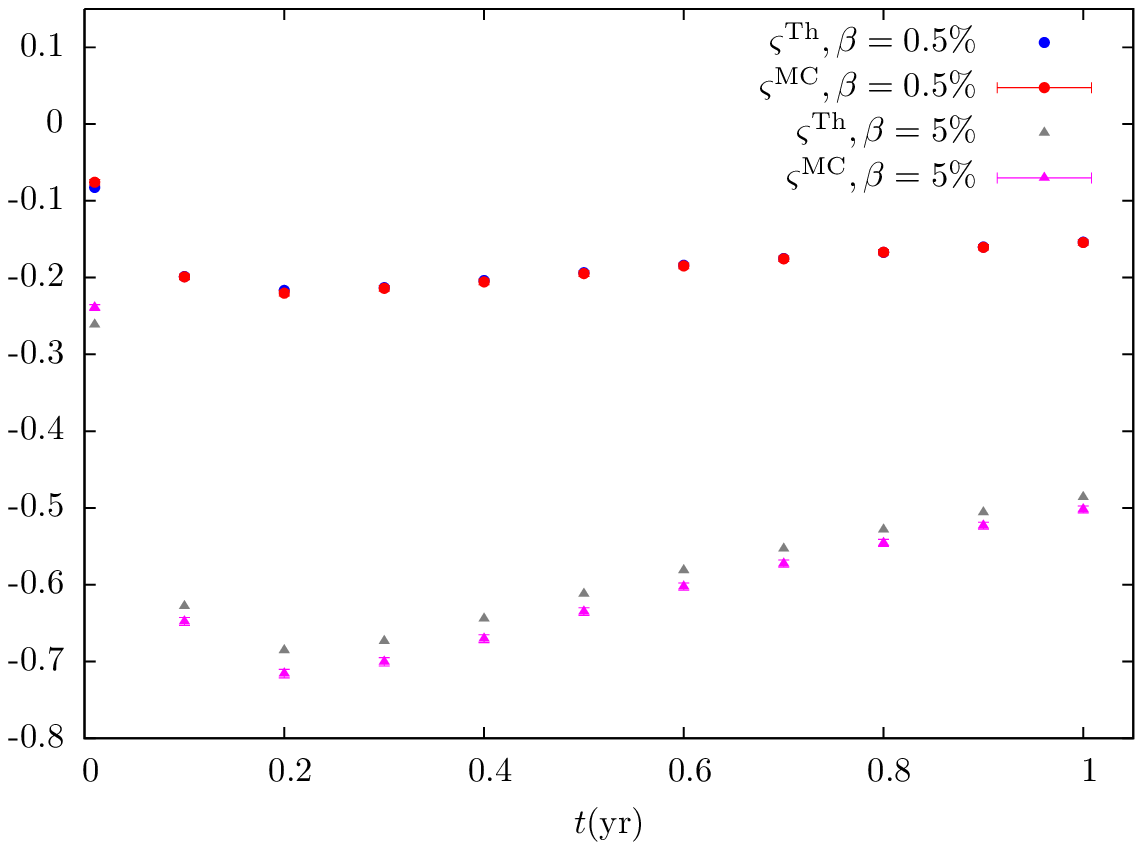}
    \end{center}
  \end{minipage}
  \begin{minipage}[b]{0.5\textwidth}
    \begin{center}
      \includegraphics[scale = 0.68]{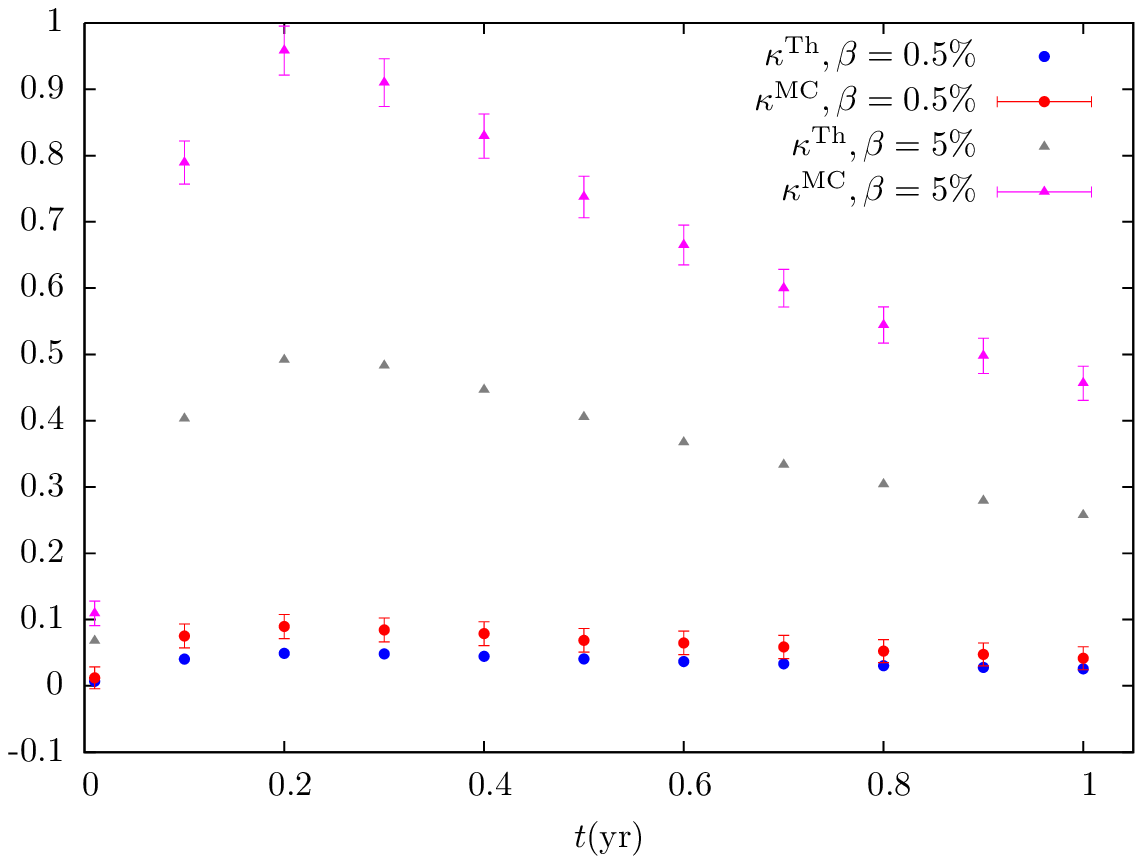}
    \end{center}
  \end{minipage}
\end{figure}

In order to make our considerations more quantitative, we present in Fig.~\ref{fig:FNRScalingRho-0.9} the scaling of
mean, variance, skewness and kurtosis with time. In each panel we report the theoretical prediction and the numerical
MC results with 95\% error bars for $\beta=0.5\%$ and $\beta=5\%$. Tab.~\ref{tab:betascaling} contains similar results
but for all the $\beta$ considered in Fig.~\ref{fig:FNRBetaRho-0.9} and $t-t_0=1$. 
\Table{\label{tab:betascaling} Scaling of normalized cumulants for increasing values of $\beta$. 
The time horizon is given by $t-t_0=1$, while the other parameters 
are the same of Fig.~\ref{fig:FNRBetaRho-0.9}. The index $^{\mathrm{MC}}$ refers to values numerically
computed with MC simulation (between parenthesis we report the error on the last significant digit, 
95\% confidence level), while cumulants with index $^{\mathrm{Th}}$ correspond to Eq.~\eref{eq:k1}-Eq.~\eref{eq:k4}.}
\br
$\beta(\%)$ & 0.5 & 1 & 2 & 5 & 10 & 25 & 50\\
\br
$k_1^{\mathrm{MC}}$&-0.00503(8)&-0.00508(8)&-0.00518(8)&-0.0055(1)&-0.0060(1)&-0.0080(1)&-0.0131(2)\\
$k_1^{\mathrm{Th}}$&-0.005&-0.005&-0.005&-0.005&-0.005&-0.005&-0.005\\
\mr
$k_2^{\mathrm{MC}}$&0.01013(1)&0.01025(1)&0.01048(1)&0.01118(2)&0.01242(2)&0.01702(4)&0.02932(8)\\
$k_2^{\mathrm{Th}}$&0.01&0.01&0.01&0.01&0.01&0.01&0.01\\
\mr
$\varsigma^{\mathrm{MC}}$&-0.154(4)&-0.219(4)&-0.311(4)&-0.502(4)&-0.733(6)&-1.29(1)&-2.22(2)\\
$\varsigma^{\mathrm{Th}}$&-0.154&-0.217&-0.307&-0.486&-0.687&-1.087&-1.54\\
\mr
$\kappa^{\mathrm{MC}}$&0.04(2)&0.08(2)&0.17(2)&0.46(2)&0.99(4)&3.2(1)&10.3(6)\\
$\kappa^{\mathrm{Th}}$& 0.026&0.0515&0.10&0.26&0.51&1.29&2.6\\
\br
\end{tabular}
\end{indented}
\end{table}

For $\beta\leq 10\%$ mean and variance computed with Eq.~\eref{eq:k1} and Eq.~\eref{eq:k2} are in statistical 
agreement with the MC estimates and the linear scaling with time theoretically 
predicted is confirmed by the numerical simulations. The compatibility of $\varsigma$ is limited to small values of $\beta$ values while the situation
is quite unsatisfactory for $\kappa$ and $\beta\geq 2\%$. All these empirical evidences strongly suggest 
that the expansion of Eq.~(\ref{eq:phiFPforward}) for $\omega_2$ nearly zero and up to
order $1/\lambda$ predicts the correct behavior of $C(\omega_1,y_0,\tau)$ for small $\beta$ values only. 
The condition $\lambda\gg 1$ is not enough to guarantee that the Edgeworth expansion \eref{eq:pXEdgeworth} 
is a good approximation of the true distribution.  
When the stationary variance of the log-volatility becomes high, the fluctuations of $Y$ sensibly deviate from $\gamma$
and the exponential function enhances the fluctuations of $X$. This mechanism is responsible for the growth 
of the empirical kurtosis at high $\beta$.  

%
%
\section{Limit case II: expansion for small $\beta$}\label{s:beta}

The numerical results shown in Section~\ref{s:lambda} suggest that the level of the stationary variance 
of $Y$ is a crucial parameter to describe accurately the returns distribution and 
for this reason we focus on it. The intuition behind our choice is that, since $\beta$ governs the level of the
fluctuations of $Y$ around its stationary value, keeping $\beta$ low allows us to linearize the exponential 
form of the volatility. Moreover we will show in the next subsection how it is possible 
to exactly solve the Fokker-Planck equation associated with the system of SDEs by means of linearization. 
We limit our investigation to $0<\beta\leq 0.1$. 
Higher values can be explored and the goodness of the approximation 
can be tested comparing the numerical results obtained via MC simulation of the linear and complete dynamics.
However, it is important to note that in practical applications it is usually required 
to keep the volatility non negative.
The probability for the volatility to become negative can be easily computed and is given by the following formula
\begin{equation}
  \frac{1}{2}\mathrm{Erfc}\left(\frac{1+\gamma+(y_0-\gamma)\ue^{-\alpha(t-t_0)}}
       {\sqrt{\beta(\ue^{-2\alpha(t-t_0)}+1)}}\right)
  \stackrel{t-t_0\gg 1/\alpha}{\longrightarrow} 
  \frac{1}{2}\mathrm{Erfc}\left(\frac{1+\gamma}{\sqrt{\beta}}\right) \ ,
\end{equation}
where Erfc is the complementary error function. After few relaxation times $1/\alpha$ and for $\gamma$ of order 0 
we have that the probability reduces to one half of $\mathrm{Erfc(\sqrt{2\alpha}/k)}$. For $\beta=1\%$ we have a 
probability of order $10^{-45}$ , while for $\beta=10\%$ it increases to $4\times 10^{-6}$, which is 
still a very small value.

\subsection{Exact solution for the linear model}

The starting point of our analysis is the linearization of Eq.~\eref{eq:dX}. 
Since when the $Y$ process termalizes the random 
variable fluctuates around $\gamma$, it is quite natural to expand the exponential 
around the stationary value. By defining
$\bar{m}\doteq m\ue^{\gamma}$ and introducing the random variable $Z\doteq Y-\gamma+1$, 
Eq.~\eref{eq:dX} and Eq.~\eref{eq:dY}
can be rewritten as
\begin{equation}\label{eq:dXlin}
  \eqalign{
    \ud X=-\frac{\bm^2}{2}(2Z-1)\ud t+\bm Z \ud W_1\ , \cr
    X(t_0)=0 \ ,
  }
\end{equation}
\begin{equation}\label{eq:dZ}
  \eqalign{
    \ud Z=\alpha(1-Z)\ud t+k\rho\ud W_1(t) + k\sqrt{1-\rho^2}\ud W_2(t) \ ,\cr
    Z(t_0)=y_0-\gamma+1 \ ,
  }
\end{equation}
To derive the analytical expression of $p_X(X(t)| X_0, Z_0)$ we will follow a strategy analogous to the one 
pioneered by Heston in \cite{Heston:1993}. For notational convenience we indicate $p_X(x|x_0,z_0)$ shortly as $p_X$.
The Fokker-Planck backward equation satisfied by $p_X$ is readily written as 
\begin{eqnarray}\label{eq:FPbackward}
  \frac{\partial}{\partial t_0}p_X &=\frac{\bm^2}{2}(2z_0-1)\frac{\partial}{\partial x_0}p_X 
  -\alpha(1-z_0)\frac{\partial}{\partial z_0}p_X\nonumber\\
  &-\frac{\bm^2}{2}z_0^2\frac{\partial^2}{\partial x_0^2}p_X
  -\rho k \bm z_0\frac{\partial^2}{\partial x_0 \partial z_0}p_X
  -\frac{k^2}{2}\frac{\partial^2}{\partial z_0^2}p_X\ .
\end{eqnarray}
Heston technique essentially reduces to the observation that, if we assume as a final 
time condition the expression $\ue^{i x \phi}$, 
Eq.~\eref{eq:FPbackward} is precisely the partial differential equation governing the 
evolution of the characteristic function 
$f(\phi;x_0,z_0)$ implicitly defined by
\begin{equation}\label{eq:pXcharacteristic}
  p_X(x|x_0,z_0)=\frac{1}{2\pi}\int_{-\infty}^{+\infty}\ue^{-i \phi x}f(\phi;x_0,z_0)\ud \phi \ .
\end{equation}
We try a solution of the form 
\begin{equation}\label{eq:cflinmodel}
  f(\phi; x_0, z_0)=\ue^{\mcA(t-t_0,\phi) +\mcB(t-t_0,\phi)z_0 +\mcC(t-t_0,\phi)z_0^2 + i\phi x_0}\ .
\end{equation}
We substitute it in Eq.~\eref{eq:FPbackward} and we get
\begin{eqnarray}
  \dot{\mcA} +\dot{\mcB}z_0 +\dot{\mcC}z_0^2 &=\frac{\bm^2}{2}(2z_0-1)i\phi 
  -\frac{\bm^2}{2}z_0^2(-\phi^2) -\alpha(1-z_0)(\mcB+2\mcC z_0) \nonumber\\
  &-\frac{k^2}{2}[(\mcB+2\mcC z_0)^2 +2\mcC] -\rho k \bm z_0 i \phi (\mcB+2\mcC z_0) \ ,
\end{eqnarray}
where the dot stays for a derivative {\it w.r.t.} $t_0$. If we collect the quadratic and linear terms, 
independent of $z_0$,
we have the ODEs satisfied by  $\mcC$, $\mcB$ and $\mcA$ respectively:
\begin{eqnarray}
  \label{eq:dotmcC}
  \dot{\mcC} &= \frac{\bm^2}{2}\phi^2 +2\alpha \mcC -2k^2\mcC^2 -\rho k \bm i \phi 2 \mcC \ ,\\
  \label{eq:dotmcB}
  \dot{\mcB} &= \bm^2 i\phi - 2\alpha\mcC +\alpha\mcB -2k^2\mcB\mcC -\rho k \bm i \phi \mcB \ ,\\
  \label{eq:dotmcA}
  \dot{\mcA} &= -\frac{\bm^2}{2}i\phi -\alpha\mcB -\frac{k^2}{2}(\mcB^2 +2\mcC) 
  \ ,
\end{eqnarray}
with final time conditions
\begin{eqnarray}
  \label{eq:initialdotmcC}
  \mcC(0,\phi) &= 0 \ ,\\
  \label{eq:initialdotmcB}
  \mcB(0,\phi) &= 0 \ ,\\
  \label{eq:initialdotmcA}
  \mcA(0,\phi) &= 0 \ .
\end{eqnarray}
Equation~\eref{eq:dotmcC} is a Riccati type ODE and once it has been solved, we insert the 
solution in Eq.~\eref{eq:dotmcB} and 
Eq.~\eref{eq:dotmcA} and we integrate them out in the usual way. The explicit 
expression for $\mcC$, $\mcB$ and $\mcA$ are 
quite involved. To improve the readability we define some auxiliary variables 
\begin{equation*}
  d\doteq 2\sqrt{k^2 \bm^2 \phi ^2+(\alpha -i k \bm \rho  \phi )^2}\ , \qquad b\doteq 2(\alpha -i k \bm\rho\phi)\ ,
\end{equation*}
\begin{equation*}
  g\doteq\frac{b-d}{b+d}\ , \qquad h\doteq i \bm^2 \phi \qquad \mathrm{and} \qquad  n\doteq\frac{\alpha}{2k^2}(b-d) \ .
\end{equation*}
Now the desired functions read
\begin{equation}\label{eq:mcC}
  \mcC(t-t_0,\phi)=\frac{b -d}{4k^2}\frac{1-\ue^{-d (t-t_0)}}{1-g\ue^{-d(t-t_0)}} \ ,
\end{equation}
\begin{eqnarray}\label{eq:mcB}
  \mcB(t-t_0,\phi)=2\frac{\ue^{-\frac{1}{2} d (t-t_0)} ((g+1) h-2 n)+n+\ue^{-d (t-t_0)} 
    (n-g h)-h}{d \left(1-g\ue^{-d (t-t_0)} \right)},
\end{eqnarray}
and
\begin{eqnarray}\label{eq:mcA}
  \mcA &(t-t_0,\phi) = \frac{1}{2} h (t-t_0) \nonumber\\
  & +2\alpha\left\{ \frac{(g+1)h -2n}{d^2\sqrt{g}}\left[ \ln(1-\sqrt{g}\ue^{-\frac{d}{2}(t-t_0)})-\ln(1-\sqrt{g})\right.\right.\nonumber\\
    &\qquad\qquad\qquad\left.-\ln(1+\sqrt{g}\ue^{-\frac{d}{2}(t-t_0)})+\ln(1+\sqrt{g})\right]\nonumber\\
  &\qquad +\frac{n(g+1)-2gh}{d^2g}\left[\ln(1-g\ue^{-d(t-t_0)})-\ln(1-g)\right]\nonumber\\
  &\qquad\left. +\frac{n-h}{d}(t-t_0)\right\}\nonumber\\
  & +2k^2\left\{ -\frac{(n-gh)^2}{d^3g}\left[ \frac{\ue^{-d(t-t_0)}-1}{(1-g)(1-g\ue^{-d(t-t_0)})} + \frac{1}{g}\Big( \ln(1-g\ue^{-d(t-t_0)})\right.\right.\nonumber\\
    &\qquad\qquad -\ln(1-g) \Big)\Big]\nonumber\\
  &\qquad -\frac{((g+1)h -2n)^2+2(n-gh)(n-h)}{d^3}\times\frac{\ue^{-d(t-t_0)}-1}{(1-g)(1-g\ue^{-d(t-t_0)})}\nonumber\\
  &\qquad +\frac{(n-h)^2}{d^2}\left[(t-t_0) +\frac{1}{d}\left( \ln(1-g\ue^{-d(t-t_0)})-\ln(1-g) \right) \right.\nonumber\\
    &\qquad\qquad\left. -\frac{g}{d}\frac{\ue^{-d(t-t_0)}-1}{(1-g)(1-g\ue^{-d(t-t_0)})}\right]\nonumber\\
  &\qquad +\left[ \frac{((g+1)h -2n)^2}{d^3g\sqrt{g}}-\frac{((g+1)h -2n)(n-h)}{d^3\sqrt{g}}\right]\nonumber\\
  &\qquad \quad\times\left[ \ln(1+\sqrt{g}\ue^{-\frac{d}{2}(t-t_0)})-\ln(1+\sqrt{g})-\ln(1-\sqrt{g}\ue^{-\frac{d}{2}(t-t_0)})\right.\nonumber\\
    &\qquad\qquad +\ln(1-\sqrt{g}) \Big]\nonumber\\
  &\qquad -2\left[ \frac{((g+1)h -2n)^2}{d^3g}+\frac{((g+1)h -2n)(n-h)}{d^3}\right]\nonumber\\
  &\qquad\quad\left.\times \frac{(1+g\ue^{-\frac{d}{2}(t-t_0)})(\ue^{-\frac{d}{2}(t-t_0)}-1)}{(1-g)(1-g\ue^{-d(t-t_0)})}\right\}\nonumber\\
  & +\frac{b-d}{4}\left\{ (t-t_0) + \frac{g-1}{dg} \left[ \ln(1-g\ue^{-d(t-t_0)})-\ln(1-g) \right]\right\}
\end{eqnarray}
The three functions $\mcA$, $\mcB$ and $\mcC$ require some care. Indeed, similarly to the
Heston case, non trivial problems emerge due to the multi-valued nature of complex square root
and logarithm. In particular, due to ``branching'' effects, the characteristic function can become
discontinuous. Following the same arguments discussed in \cite{Albrecher:2007,Kahl:2005,Lord:2006}, we 
checked the smoothness of $f(\phi;x_0,z_0)$ for the set of parameters used in the next Section.

The complete characterization of $f(\phi;x_0,z_0)$ is the main result of this paper. The only remaining 
task to be performed to obtain the probability distribution in the direct space is to Fourier anti-transform 
it. We stress again that the obtained solution is an exact one but for the linear model. 
In the following Section we provide a set of $\beta$ values for which we have numerically tested the 
effectiveness of the approximation.

\subsection{Further Numerical Results}

The exact solution for the probability distribution of returns in the framework of the linear 
problem is given in an implicit way by means of a Fourier anti-transformation of Eq.~\eref{eq:pXcharacteristic}, 
which can not be computed analytically. We compared two different ways for the numerical evaluation of $p_X$: 
the integration with a trapezoidal approximation and the FFT algorithm. In order to 
improve the efficiency of the methods one can observe that $p_X$ is a real function, so the integrand of
 Eq.~\eref{eq:pXcharacteristic}  must have its real part even and its imaginary part odd. We can integrate 
on one half of the entire real axis and take twice the real part of the result. To sum up:
\begin{eqnarray}
   p_X(x|x_0,z_0)=\frac{1}{\pi}\mathrm{Re}\left[\int_{0}^{+\infty}\ue^{-i \phi x}f(\phi;x_0,z_0)\ud \phi\right]\ .
\end{eqnarray}
\begin{figure}[t!]
  \caption{\label{fig:SNRBetaRho-0.9} Returns distributions from MC simulation of exponential 
    and linear models in comparison with the probability distributions given 
    by Eq.~\eref{eq:pXcharacteristic}, numerically 
    computed by FFT algorithm. Parameters values as 
    discussed in the text, curves from bottom to top correspond to increasing values of $\beta$: 0.5\%,
    1\%, 2\%, 5\% and 10\%. 
    Curves have been shifted upwards.
    Top panel: $\rho=0.5$ and $t-t_0=1$; bottom panel: $\rho=-0.9$ and $t-t_0=1$.
  }
  \begin{center}
    \includegraphics[scale = 0.85]{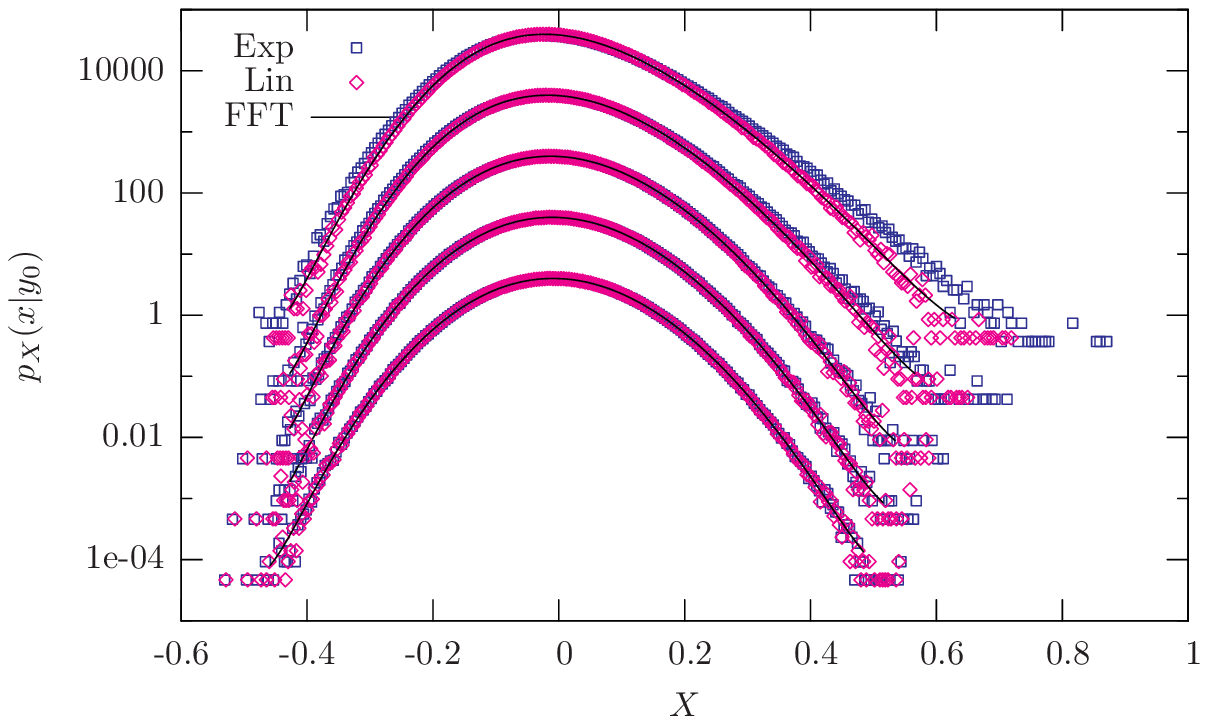}
  \end{center}
  \begin{center}
    \includegraphics[scale = 0.85]{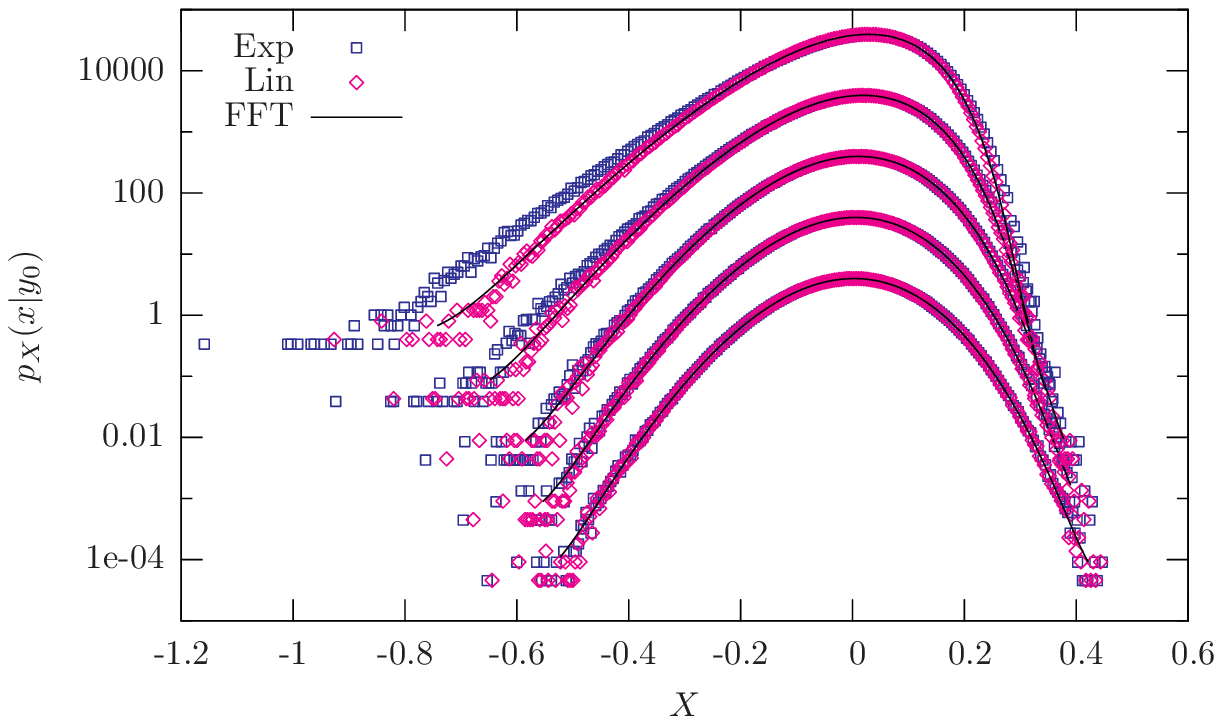}
  \end{center}
\end{figure}

We checked the convergence of the two methods mentioned previously increasing the number of points for the 
sampling of the integrand. All the results presented in the next figures have been obtained 
with the FFT algorithm with $2^{22}$ points on the fixed interval of integration $[0,10^3]$. The tails 
of the resulting distribution must decrease to zero, but they are extremely sensitive to the FFT frequency sampling  
so they have been neglected in Fig.~\ref{fig:SNRBetaRho-0.9}. For the reasons explained in Section~\ref{s:FNR}, we 
chose $\Delta t=10^{-4}$ and MCPATHS$=5 \cdot 10^6$ to simulate the dynamics of the exponential and linear models.
\begin{figure}[ht!]
  \caption{\label{fig:SNRscaling} From top left clockwise: scaling with
    time of mean, variance, kurtosis and skewness. 
    Comparison between numerically estimated values of normalized cumulants for the exponential 
    (Eq.~\eref{eq:dY}, Eq.~\eref{eq:dX}) and the linear models (with MC 95\% confidence level) . 
    $\rho=-0.9$, $\beta=0.5\%, 5\%$ and other parameters as in the text.
  }
  \begin{minipage}[b]{0.5\textwidth} 
    \begin{center}
      \includegraphics[scale = 0.68]{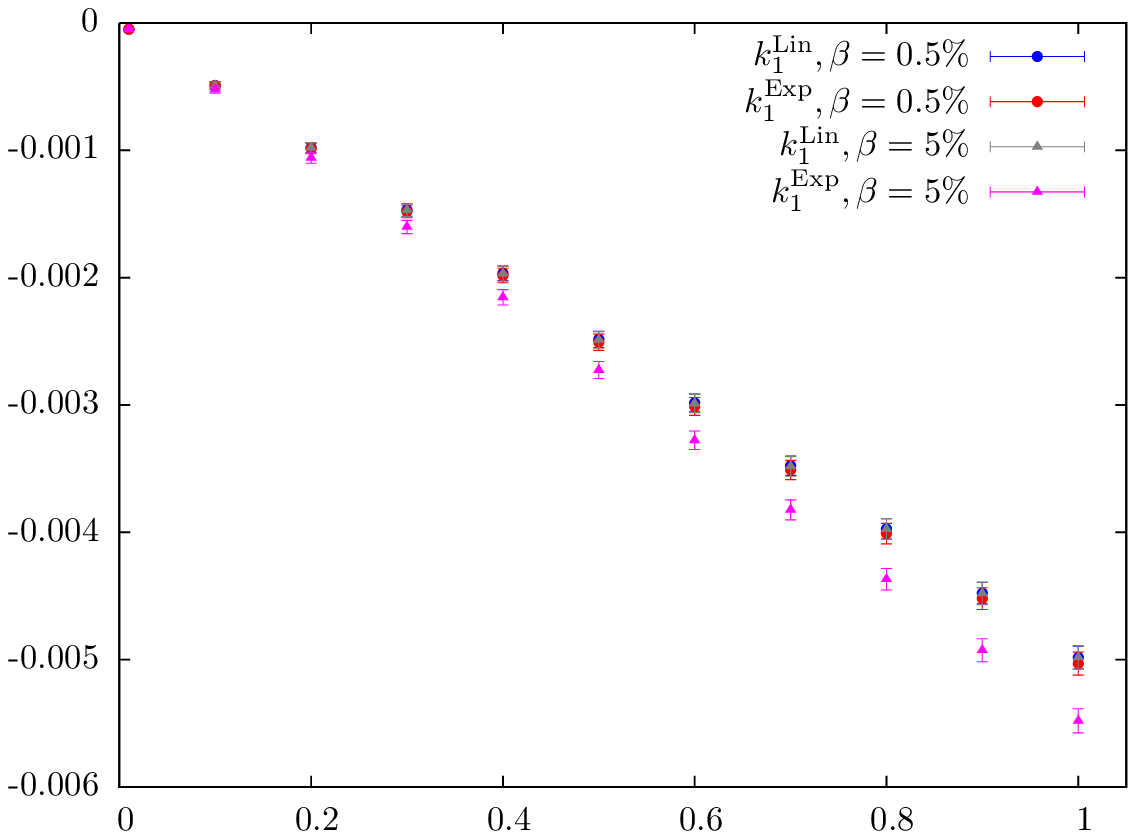}
    \end{center}
  \end{minipage}
  \begin{minipage}[b]{0.5\textwidth}
    \begin{center}
      \includegraphics[scale = 0.68]{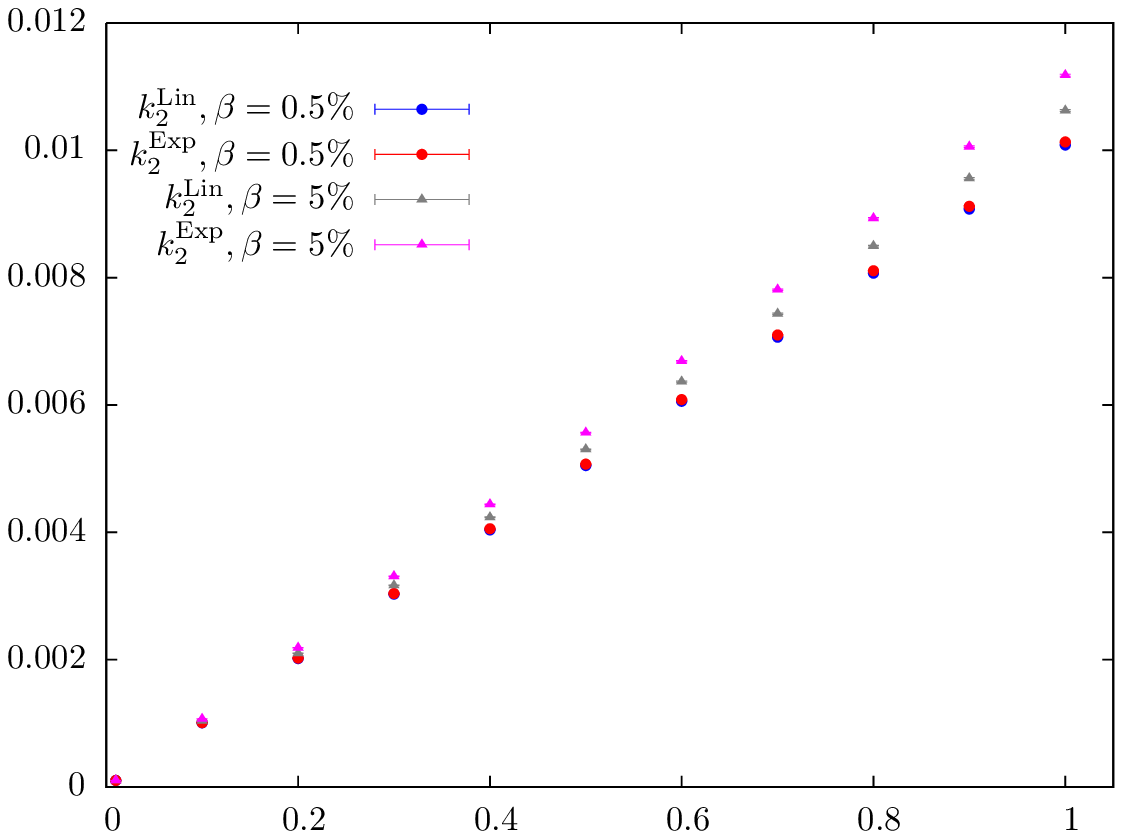}
    \end{center}
  \end{minipage}
  \begin{minipage}[b]{0.5\textwidth}
    \begin{center}
      \includegraphics[scale = 0.68]{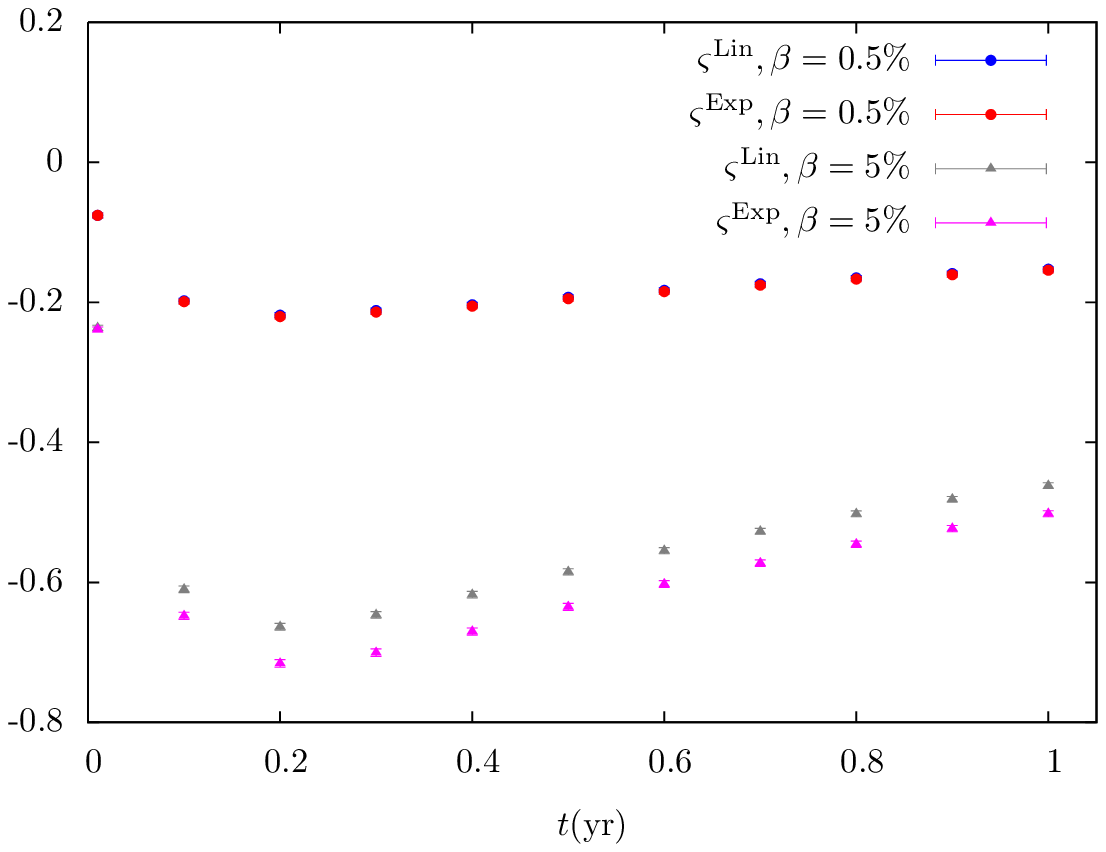}
    \end{center}
  \end{minipage}
  \begin{minipage}[b]{0.5\textwidth}
    \begin{center}
      \includegraphics[scale = 0.68]{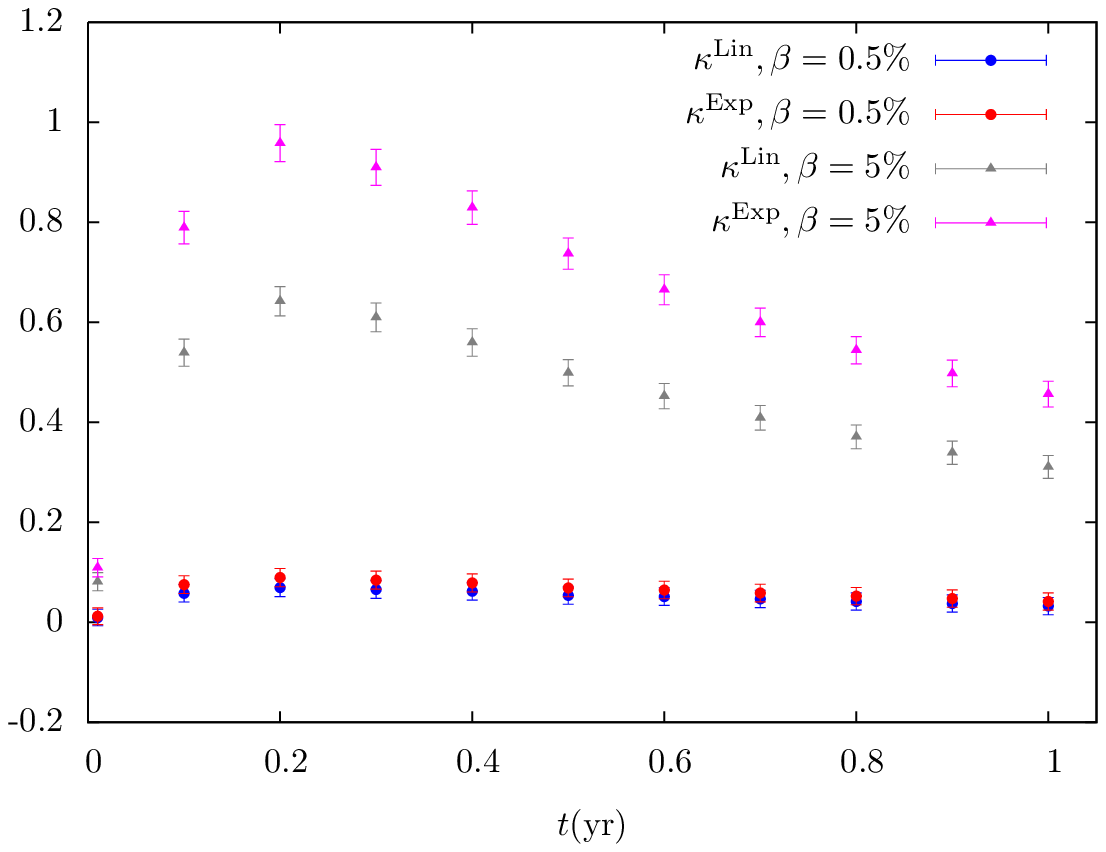}
    \end{center}
  \end{minipage}
\end{figure}

In Fig.~\ref{fig:SNRBetaRho-0.9} we report the histograms of returns obtained via MC simulations of the exponential 
and the linear models with the $p_X$ curve given by the numerical Fourier anti-transformation. 
$m=0.1$, $\alpha=10$, $\gamma=0$ and $y_0=0$ are fixed for all the figures and table of this Section. 
Each group of three curves corresponds to $\beta$ values 0.5\%, 1\%, 2\%, 5\% and 10\%, from bottom to top respectively.
Again each curve has been shifted for the sake of readability. 
The top frame has been obtained with a positive $\rho$ value 
(0.5), which is evident from the rightward asymmetry of the curves, while, in the bottom panel, curves have large
leftward asymmetry corresponding to $\rho=-0.9$. 
The agreement between theory prediction and MC simulation is very good for $\beta=0.5\%$, 1\% and 2\%, in 
both cases of $\rho$ values. The distributions of returns from exponential model are well reproduced by the ones 
from linear model and, as a consequence, by the semi-analytical approximation $p_X$. 
The agreement worsen when $\beta$ increases to $5\%$ and $10\%$, as it can be seen observing the fatter tail in 
each panel. All the curves in Fig.~\ref{fig:SNRBetaRho-0.9}
show that the returns distribution in the linear model is exactly the one computed via FFT, thus confirming 
the goodness of the exact solution. Finally we observe the good behavior of all the
semi-analytical curves, which, since they are exact solutions of a Fokker-Planck equation, do not become negative
or fluctuate like in the case of the Edgeworth expansion considered in the previous Section.

Again, to make the considerations more quantitative, in Fig.~\ref{fig:SNRscaling} the scaling with time of mean, 
variance, skewness and kurtosis is shown for $\beta=0.5\%$ and 5\%, $\rho=-0.9$. 
Then in Tab.~\ref{tab:cumulantscaling}, the 
numerical values of normalized cumulants computed with MC samples of returns 
from exponential and linear models are shown 
for $\beta=1\%, 2\%$ and $t-t_0=0.01,0.1,0.2,0.5,1$. The results show that, for $\beta \leq 2\%$, mean, skewness and 
kurtosis from the complete and linear models agree within the usual 95\% MC confidence level, 
while variance values are not statistically compatible. 
This is an expected result, since the linear model can not exactly reproduce the complete one and for this reason
we also expect that for higher values of MCPATHS the discrepancies for the other cumulants should emerge. 
However, we are able to evaluate the relative differences among cumulants values and we can evaluate the degree 
of approximation of the analytical solution: for $\beta=1\%$ the relative disagreement between $k_2^{\mathrm{Exp}}$ 
and $k_2^{\mathrm{Lin}}$ increases with time from 0.6\% to 1\%.
As expected, the disagreement between cumulants computed for the exponential and linear models increases with 
the value of $\beta$.

In comparison with the results given in Section~\ref{s:FNR}, one can note that the tails of the distribution are
better reproduced by the semi-analytical approximation (for mentioned values of $\beta$) than by the Edgeworth 
expansion, the other parameters being fixed. Also the general agreement of cumulants for $\beta \leq 2\%$
is better than the corresponding results given in Fig.~\ref{fig:FNRScalingRho-0.9} and Tab.~\ref{tab:betascaling}.
\Table{\label{tab:cumulantscaling} Scaling with time of normalized cumulants for $\beta=1\%$ and $2\%$, 
  $\rho=-0.9$ and other parameters as in the text. The indexes $^{\mathrm{Exp}}$ and $^{\mathrm{Lin}}$ 
  refer to values numerically computed with MC simulations with exponential and linear dynamics 
  (between parenthesis we report the error on the last significant digit, 
  95\% confidence level).
}
\br
& \centre{5}{$\beta = 1\%$}\\
& \crule{5}\\
$t-t_0(\mathrm{yr})$ &0.01 & 0.1 & 0.2 & 0.5 & 1\\
\br
$k_1^{\mathrm{Exp}} (10^{-4})$ &-0.50(8)	&-6.0(2)	&-9.9(4)	&-25.3(6)	&-50.8(9)\\
$k_1^{\mathrm{Lin}}~ (10^{-4})$ &-0.50(8)	&-4.9(2)	&-9.8(4)	&-24.9(6)	&-49.8(9)\\
\mr
$k_2^{\mathrm{Exp}} (10^{-4})$ &1.002(1)	&10.15(1)	&20.42(2)	&51.26(6)	&102.5(1)\\
$k_2^{\mathrm{Lin}}~ (10^{-4})$ &1.001(1)	&10.09(1)	&20.26(2)	&50.79(6)	&101.5(1)\\
\mr
$\varsigma^{\mathrm{Exp}}$ &-0.107(3)	&-0.282(4)	&-0.312(4)	&-0.276(4)	&-0.219(4)\\
$\varsigma^{\mathrm{Lin}}$ &-0.107(3)	&-0.279(4)	&-0.306(4)	&-0.271(4)	&-0.215(4)\\
\mr
$\kappa^{\mathrm{Exp}}$ &0.02(1)	&0.15(2)	&0.18(2)	&0.14(2)	&0.08(2)\\
$\kappa^{\mathrm{Lin}}$ &0.02(2)	&0.11(2)	&0.14(2)	&0.11(2)	&0.07(2)\\
\br
& \centre{5}{$\beta = 2\%$}\\
& \crule{5}\\
$k_1^{\mathrm{Exp}} (10^{-4})$
&-0.50(8)        &-5.0(2)        &-10.1(4)        &-25.8(6)        &-51.8(9)\\
$k_1^{\mathrm{Lin}}~ (10^{-4})$
&-0.50(8)        &-4.9(2)        &-9.8(4)        &-24.9(6)         &-49.8(9)\\
\mr
$k_2^{\mathrm{Exp}} (10^{-4})$
&1.004(1)        &10.28(1)        &20.77(2)        &52.35(7)        &104.8(1)\\
$k_2^{\mathrm{Lin}}~ (10^{-4})$ 
&1.002(1)        &10.16(1)        &20.45(2)        &51.38(6)        &102.8(1)\\
\mr
$\varsigma^{\mathrm{Exp}}$ &-0.151(4)        &-0.402(4)        &-0.443(4)        &-0.393(4)        &-0.311(4)\\
$\varsigma^{\mathrm{Lin}}$ &-0.150(4)        &-0.392(4)        &-0.429(4)        &-0.380(2)        &-0.300(4)\\
\mr
$\kappa^{\mathrm{Exp}}$ &0.04(2)        &0.30(2)        &0.36(2)        &0.28(2)        &0.17(2)\\
$\kappa^{\mathrm{Lin}}$ &0.03(2)        &0.22(2)        &0.27(2)        &0.21(2)        &0.13(2)\\
\br
\end{tabular}
\end{indented}
\end{table}
%

%
%
\section{Real data analysis}\label{s:realdata} 

We test the effectiveness of previous analytical results on a data set composed by 
several financial indexes. In particular we consider the following indexes from the equity sector, 
DAX30, CAC40, FTSE100, S\&PMib, S\&P500, DJ Euro Stoxx 50 and NYSE, but  
we detail the analysis only for DAX30 and DJ Euro Stoxx 50. 
Similar results have been found also for the other series.\par
DAX30 time series is made of $12173$ daily close prices, from $4^{th}$ January 1960 until $30^{th}$ June 2008, while 
for DJ Euro Stoxx 50 we consider $5566$ prices, from $1^{st}$ January 1987 until $31^{st}$ August 2008. 
Time is measured on a yearly base, so for DAX30 and DJ Euro Stoxx 50 we have 48.5 years and 21.75 years time windows 
respectively. The model described by Eqs.~\eref{eq:dS} and \eref{eq:dY} depends on 8 parameters. However, 
$s_0$ corresponds to the asset spot price so the true free parameters are $\mu$, $m$, $y_0$, $\alpha$, $\gamma$, 
$\beta$ and $\rho$. In order to estimate their values we adopt the following strategies:
\begin{enumerate}
  \item[$\mu$:] from the discretized version of equation (\ref{eq:dS}) 
    \begin{equation}
      \frac{\Delta S_i}{S_i} = \mu\,\Delta t + m\,\sqrt{\Delta t}\,\ue^{Y_i}\epsilon_i,
    \end{equation}
    where $\Delta S_i=S_{i+1}-S_i$ and $\epsilon_i\sim\mathcal{N}(0,1)$, we can conclude that 
    the expectation $\langle \Delta S_i/S_i\rangle/\Delta t$ over the real data sample
    provides an estimate of $\mu$. For DAX30 $\Delta t=3.98\times 10^{-3}$, while for 
    DJ Euro Stoxx 50 we have $\Delta t=3.91\times 10^{-3}$.
    \item[$\gamma$, $y_0$:] Remembering the definition $\sigma(t)=m\ue^{Y(t)}$, it is readily proved that 
      \begin{equation}
	\uE{\sigma(t)^n}=m^n\,\ue^{n\uE{Y(t)}}\ue^{\frac{n^2}{2}\left(\uE{Y(t)^2}-\uE{Y(t)}^2\right)},
      \end{equation}
      with $\uE{Y(t)}$ and $\uE{Y(t)^2}$ given in Eqs.~\eref{eq:Ymean} and \eref{eq:Yvar}.
      It is common practice to assume that, for the observed time series, the volatility process has already reached
      the stationary state. Under this assumption, previous expression reduces to 
      $\uE{\sigma(t)^n}= \left(m \ue^{\gamma}\right)^n \ue^{\frac{n^2}{2}\beta}$. It is crucial noticing that
      all the moments do not depend on $y_0$ and $\ue^{\gamma}$ is always coupled with $m$. 
      For this reason we introduce the parameter $\bar{m}=m\ue^{\gamma}$ 
      (see also lines before Eq.\eref{eq:dX}) and we set $y_0$ equal to zero.
      \item[$\bar{m}$,$\beta$:] These two parameters completely specify the stationary log-normal distribution
	of the stochastic variable $\sigma$. To extract the distribution of the hidden variable $\sigma$ 
	from the series of financial returns, we implement the methodology described in 
	Appendix B of reference \cite{Pasquini_Serva:2000}. 
	The consistency of our code has been tested over the NYSE time series and we have found results in full
	agreement with those quoted in \cite{Pasquini_Serva:2000}.       
      \item[$\alpha$,$\rho$:] Finally, to estimate $\alpha$ and $\rho$, we search for values able to reproduce
	the empirical scaling with time of real data skewness and kurtosis. We consider time horizons from
	one day to one hundred days and normalized cumulants are evaluated with standard estimators.
	By means of centered returns, computed from market prices, we obtain
	the empirical skewness, $\varsigma_{Ph}$, and kurtosis, $\kappa_{Ph}$, and corresponding errors, 
	$\epsilon_{Ph}^{\varsigma}$ and $\epsilon_{Ph}^{\kappa}$. By generating 10000 paths 
	from the exponential Ornstein-Uhlenbeck dynamics, we can compute the MC estimators 
	$\varsigma_{MC}$, $\kappa_{MC}$ and associated errors, 	$\epsilon_{MC}^{\varsigma}$ 
	and $\epsilon_{MC}^{\kappa}$. The optimal $\alpha$ and $\rho$ 
	are given by those values that minimize the sum over 100 time horizons of the normalized 
	squared differences, according to the following formula:
	\begin{equation}\label{eq:chisquared}
	  (\alpha^*,\rho^*)=\min_{\alpha>0,\,\rho\in(-1,1)} 
	  \sum_{i=1}^{100} \left[\frac{(\varsigma_{i,Ph}-\varsigma_{i,MC})^2}
	    {{\epsilon_{i,Ph}^{\varsigma}}^2+{\epsilon_{i,MC}^{\varsigma}}^2}+
	    \frac{(\kappa_{i,Ph}-\kappa_{i,MC})^2}
		 {{\epsilon_{i,Ph}^{\kappa}}^2+{\epsilon_{i,MC}^{\kappa}}^2}
	      \right].
	\end{equation}
	The subscript $i$ means that the corresponding quantity is evaluated at time $i\Delta t$.
	Since the presence of noisy denominators in Eq.~\eref{eq:chisquared}, 
	we can not resort to optimization algorithms based on gradient methods. 
	For this reason we implement the principal axis approach with the one dimensional search
	based on the Brent method (see \verb|opt/| directory at http://www.netlib.org). 
\end{enumerate} 
\begin{figure}[ht!]
  \caption{\label{fig:StationarySigma} Probability distribution $P(\sigma)$ of volatility for DAX30 (blue boxes)
    and DJ Euro Stoxx 50 (violet diamonds). Solid lines correspond to a log-normal fit. 
  }
    \begin{center}
      \includegraphics[scale = 0.85]{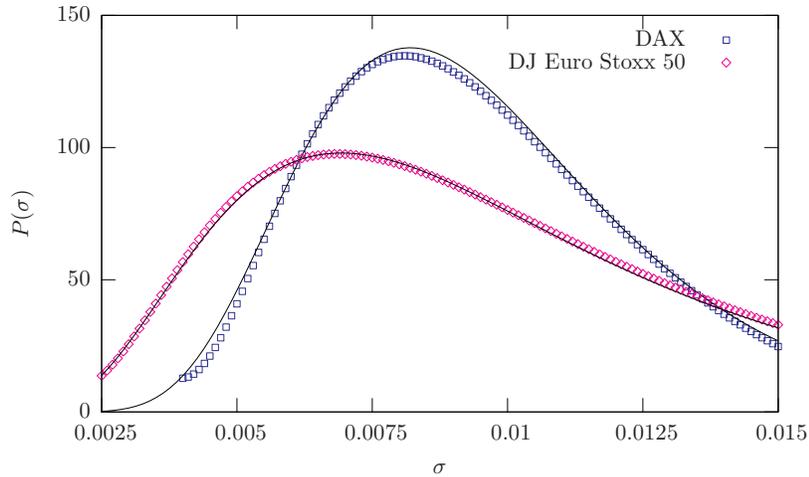}
    \end{center}
\end{figure}
The probability distribution $P(\sigma)$ of volatility is plotted in Fig.~\ref{fig:StationarySigma}, 
both for the DAX30 index and the DJ Euro Stoxx 50. Distributions tails suffer low statistics effects, 
but in the central region they are well fitted by a log-normal distribution
\begin{equation}
  p(\sigma)=\frac{1}{\sqrt{2\pi}s\sigma}\exp -\frac{1}{2}\left(
  \frac{\log\sigma-\log\sigma_0}{s}\right)^2.
\end{equation}
The fit is performed in the range $0.0004\leq\sigma\leq 0.015$ for DAX30 and gives 
$\log\sigma_0=-4.492\pm 0.001$ and $s=0.334\pm 0.001$, while for the 
DJ Euro Stoxx 50 the log-normal fit is consistent in a wider region $0.0015\leq\sigma\leq 0.015$ and
gives $\log\sigma_0=-4.7049\pm 0.0008$ and $s=0.5147\pm 0.0007$. 
Tab.~\ref{tab:DAXandSX5E} details the corresponding values for $\bar{m}=\sigma_0/\sqrt{\Delta t}$ 
and $\beta=s^2$.
\Table{\label{tab:DAXandSX5E} Estimated values for the model parameters for DAX30 and DJ Euro Stoxx 50 indexes.}
\br
& $\mu\,(\mathrm{yr}^{-1})$ & $y_0$ & $\gamma$ & $\bar{m}\,(\mathrm{yr}^{-\frac{1}{2}})$ 
& $\beta$ & $\alpha\,(\mathrm{yr}^{-1})$ & $\rho$\\
\br
DAX30 & $7.39\times 10^{-2}$ & 0 & 0 & $14.52\times 10^{-2}$ & $11.16\times 10^{-2}$ & 30.76 & -0.54 \\

\begin{figure}[ht!]
  \caption{\label{fig:DJScaling}  DJ Euro Stoxx 50 index returns distributions two different time horizons, 
    25 trading days (first row) and 45 trading days (second row).
  }
  \begin{minipage}[b]{0.55\textwidth} 
    \begin{center}
      \includegraphics[scale = 0.68]{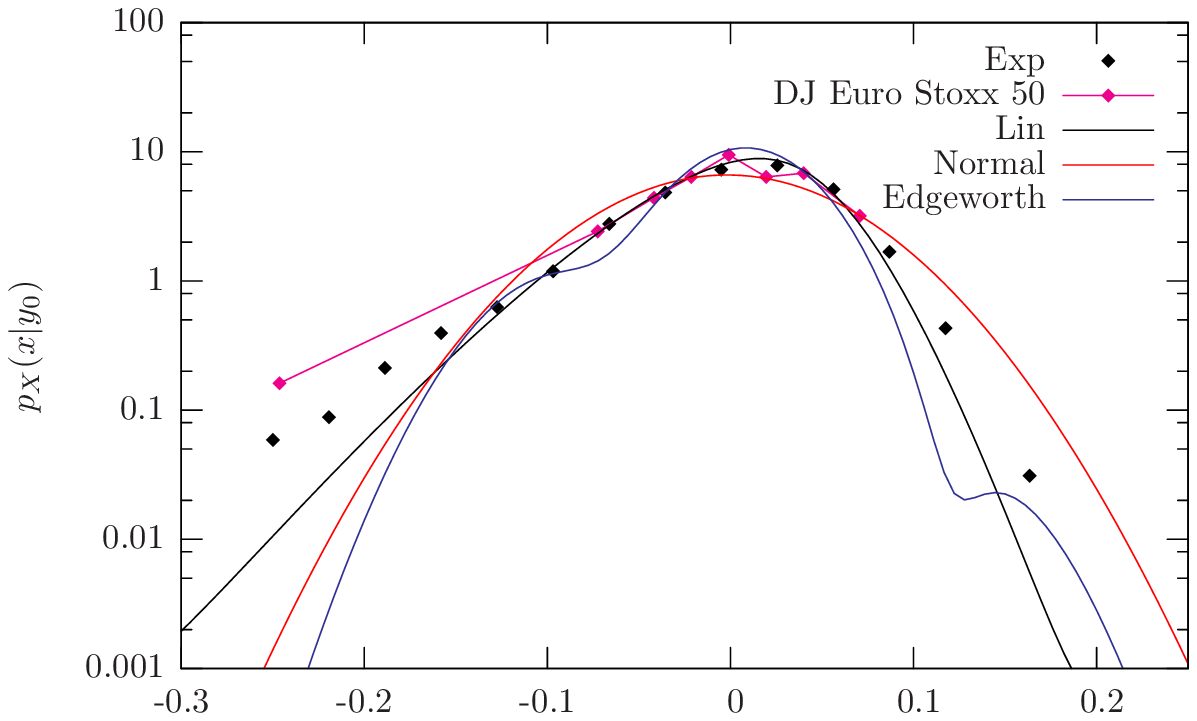}
    \end{center}
  \end{minipage}
  \begin{minipage}[b]{0.45\textwidth}
    \begin{center}
      \includegraphics[scale = 0.68]{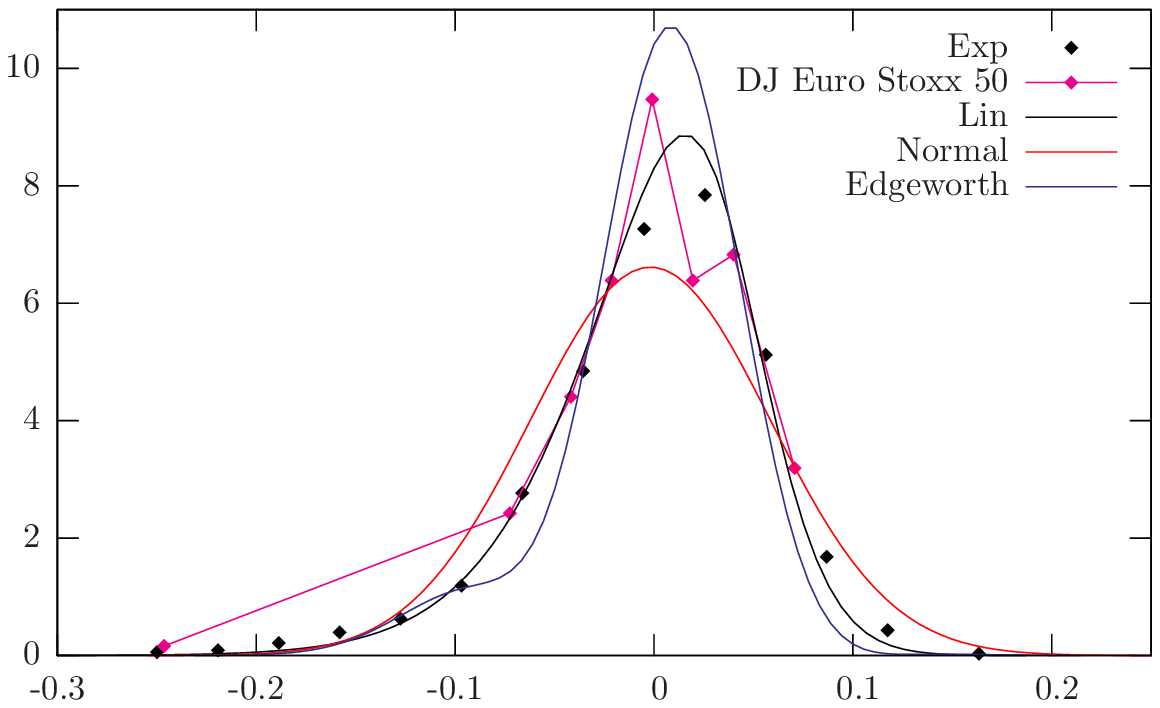}
    \end{center}
  \end{minipage}
  \begin{minipage}[b]{0.55\textwidth} 
    \begin{center}
      \includegraphics[scale = 0.68]{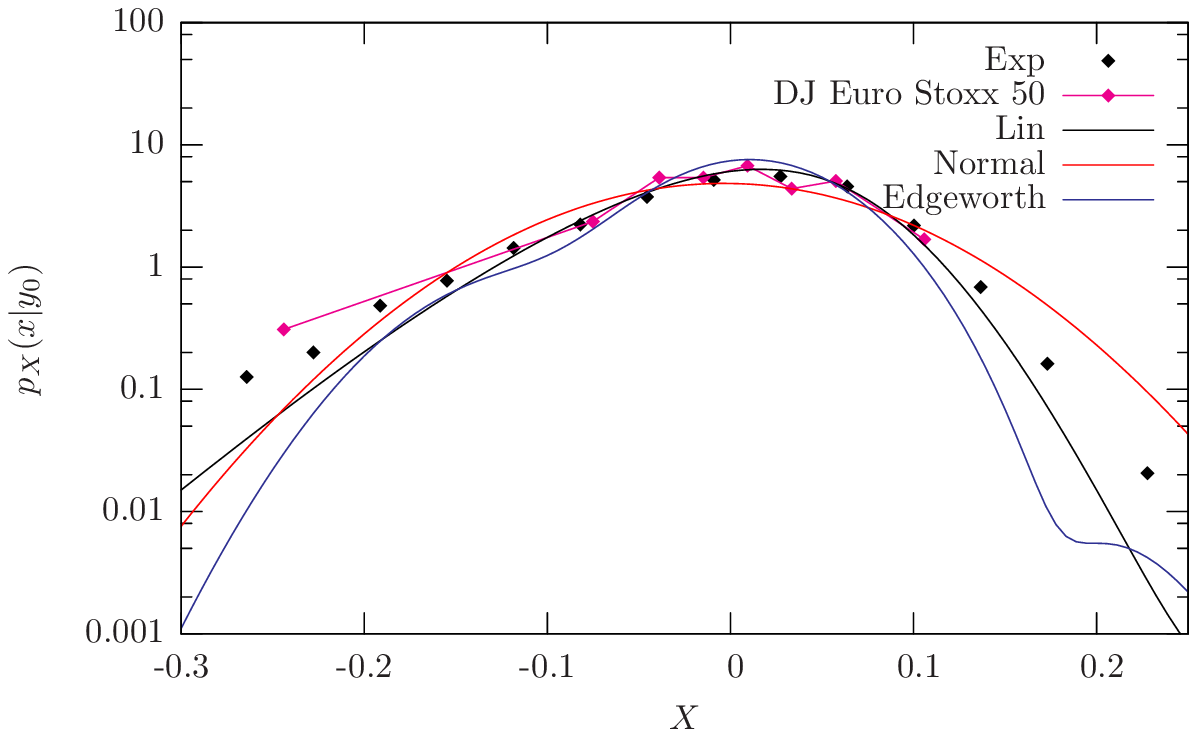}
    \end{center}
  \end{minipage}
  \begin{minipage}[b]{0.45\textwidth}
    \begin{center}
      \includegraphics[scale = 0.68]{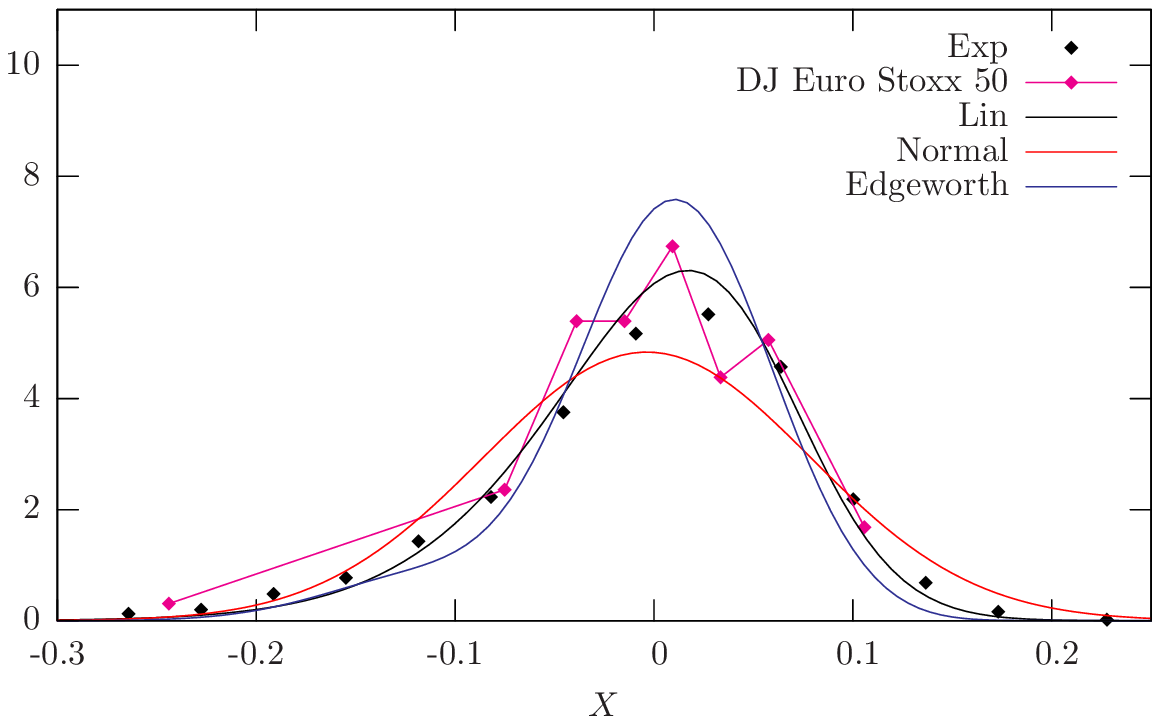}
    \end{center}
  \end{minipage}
\end{figure}

\mr
DJ Euro Stoxx 50 & $7.97\times 10^{-2}$ & 0 & 0 & $14.45\times 10^{-2}$ & $26.51\times 10^{-2}$ & 47.74 & -0.52 \\
\br
\end{tabular}
\end{indented}
\end{table}
\begin{figure}[ht!]
  \caption{\label{fig:DJNrmcumulants} Scaling with time of skewness (left panel) and kurtosis (right panel) for
    DJ Euro Stoxx 50 index. We plot the scaling of empirical cumulants (violet crossed-line) with
    the corresponding 68\% CL bars and cumulants with 68\% CL bars
    from the MC simulation (blue boxed-line). 
    The solid line corresponds to the analytical expressions given in Eqs.~\eref{eq:k3} and \eref{eq:k4}.
  }
  \begin{minipage}[b]{0.5\textwidth} 
    \begin{center}
      \includegraphics[scale = 0.68]{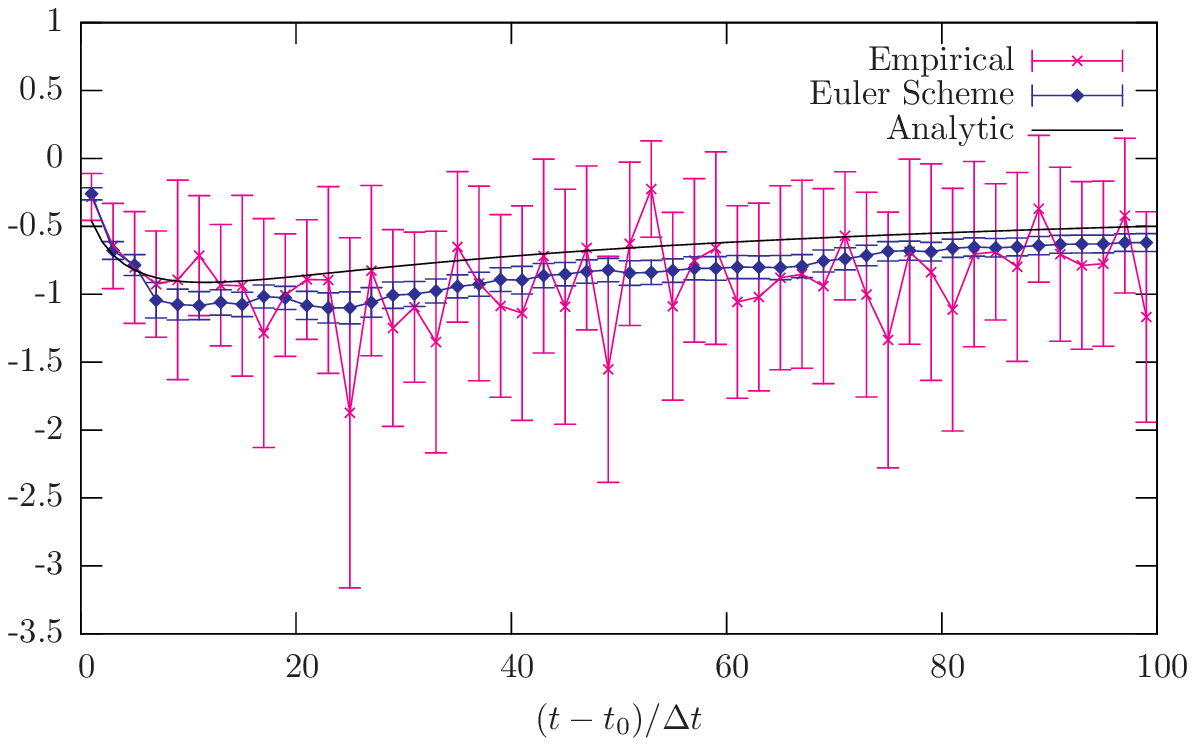}
    \end{center}
  \end{minipage}
  \begin{minipage}[b]{0.5\textwidth}
    \begin{center}
      \includegraphics[scale = 0.68]{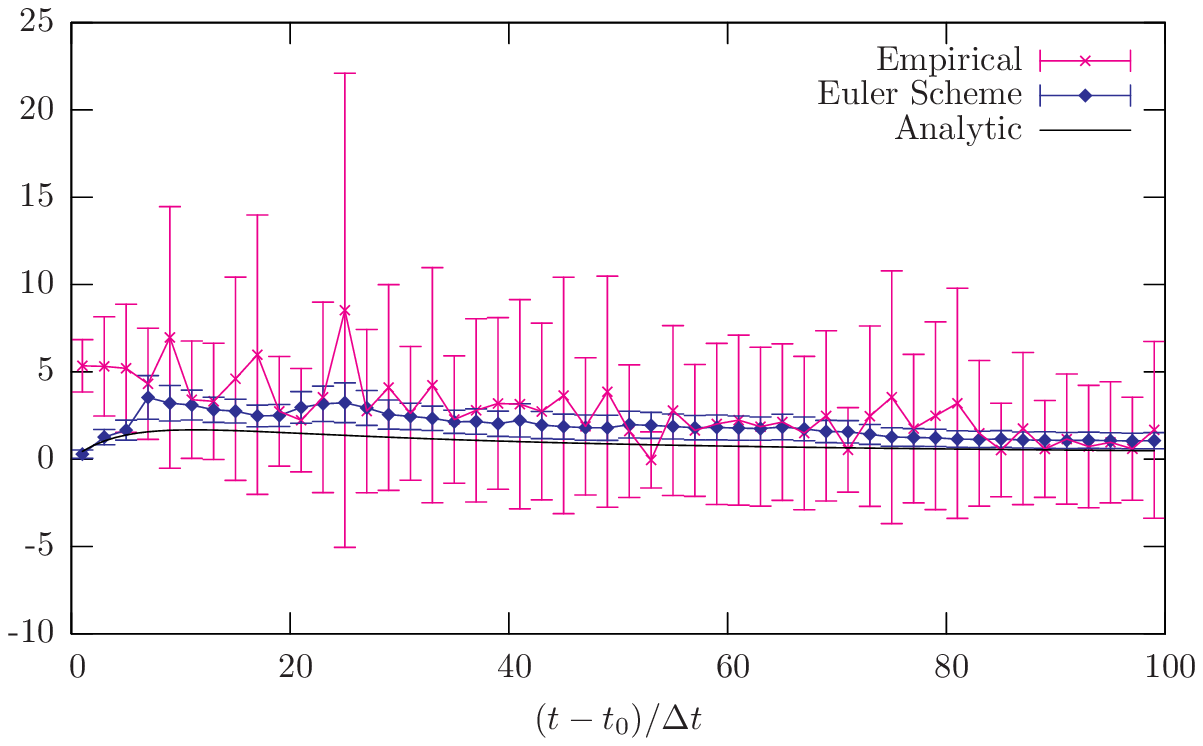}
    \end{center}
  \end{minipage}
\end{figure}
In Fig.~\ref{fig:DJNrmcumulants} we plot the scaling with time of skewness and kurtosis for the DJ Euro Stoxx 50 index. 
We have found similar results for DAX30, but we do not report them here. Error bars represent
the 68\% confidence level (CL), while the solid lines correspond to the analytical expressions given by 
Eqs.~\eref{eq:k3} and \eref{eq:k4}. These lines have been generated using parameters values obtained by the minimization
of r.h.s. of Eq.~\eref{eq:chisquared}. 
Their values are detailed in Tab.~\ref{tab:DAXandSX5E}. It is worth to comment that in order to capture the leftward 
asymmetry of the real data distribution the correlation coefficient $\rho$ is correctly predicted to be negative. 
The relaxation time $1/\alpha\sim 5$ days implies a quite fast thermalization process.  
The plotted results indicate that the exponential Ornstein-Uhlenbeck model is unable to capture 
the excess of kurtosis observed in many high-frequency returns distributions 
\cite{Bouchaud_Potters:2000,Mantegna:2000}, but it can provide reliable predictions 
for returns distributions corresponding to sufficiently long-time lags and is able
to account for the transition from a leptokurtic to a Gaussian regime, as observed
in other SV models \cite{Dragulescu_Yakovenko:2002,Masoliver_Perello:2002}.
\begin{figure}[ht!]
  \caption{\label{fig:DAXScaling} DAX30 index returns distributions for two different time horizons, 
    25 trading days (first row) and 45 trading days (second row).
  }
  \begin{minipage}[b]{0.55\textwidth} 
    \begin{center}
      \includegraphics[scale = 0.68]{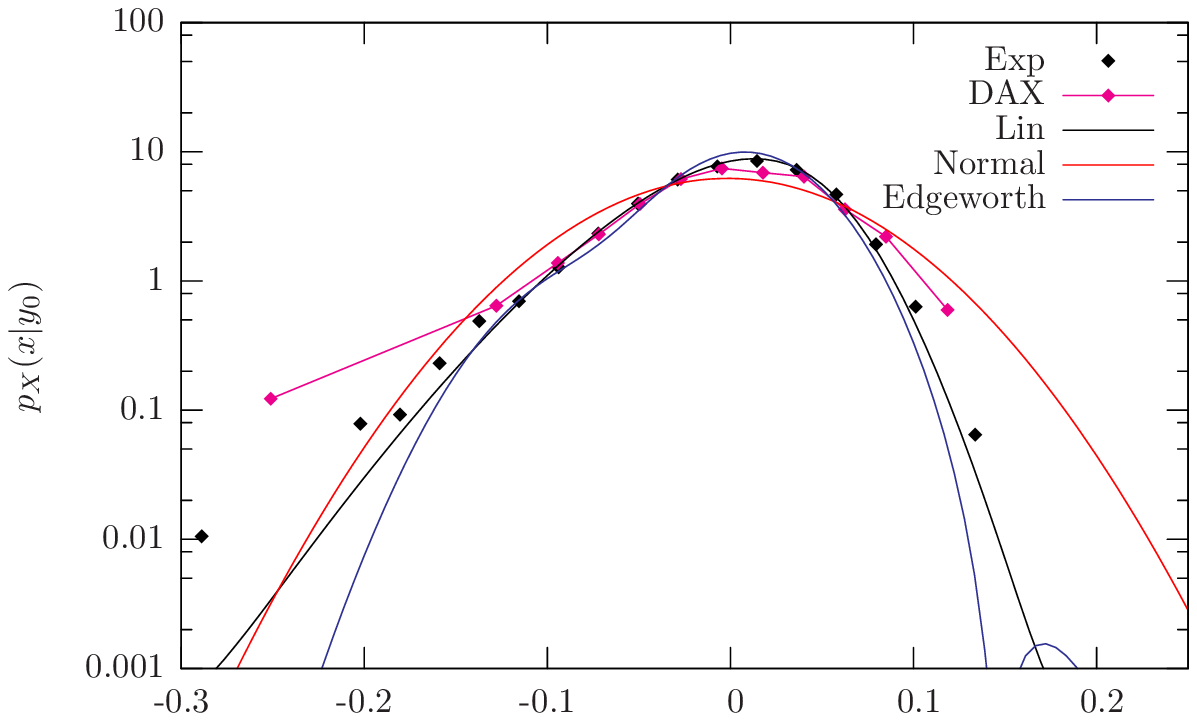}
    \end{center}
  \end{minipage}
  \begin{minipage}[b]{0.45\textwidth}
    \begin{center}
      \includegraphics[scale = 0.68]{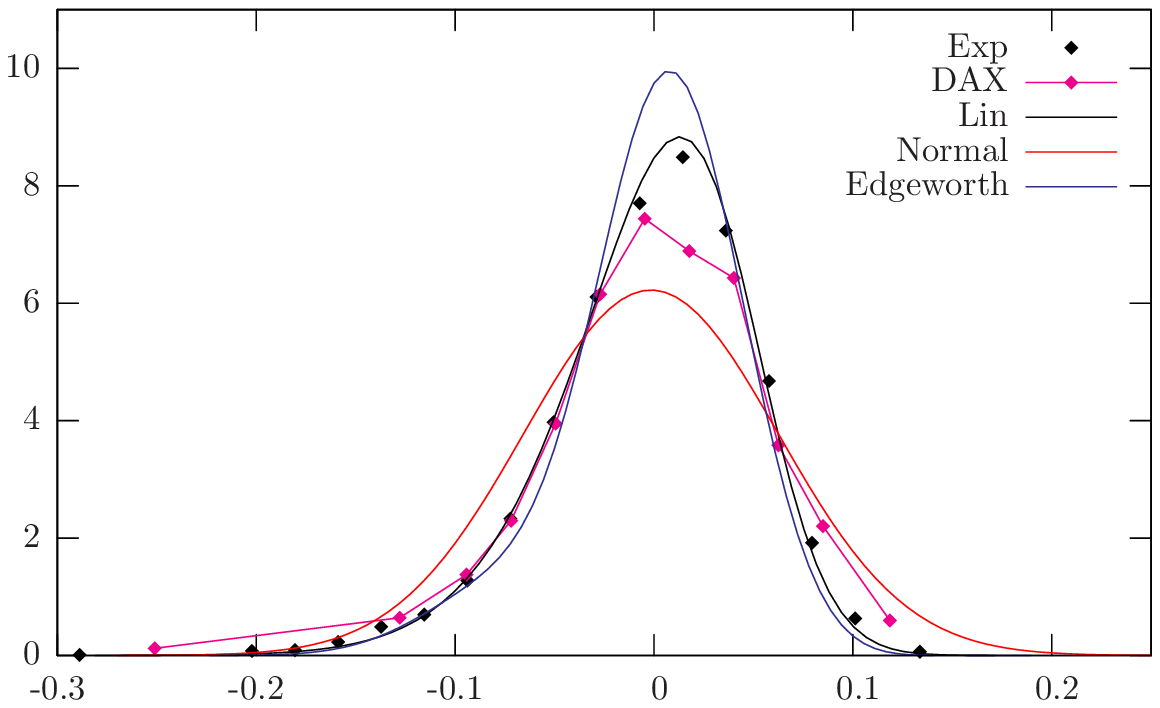}
    \end{center}
  \end{minipage}
  \begin{minipage}[b]{0.55\textwidth} 
    \begin{center}
      \includegraphics[scale = 0.68]{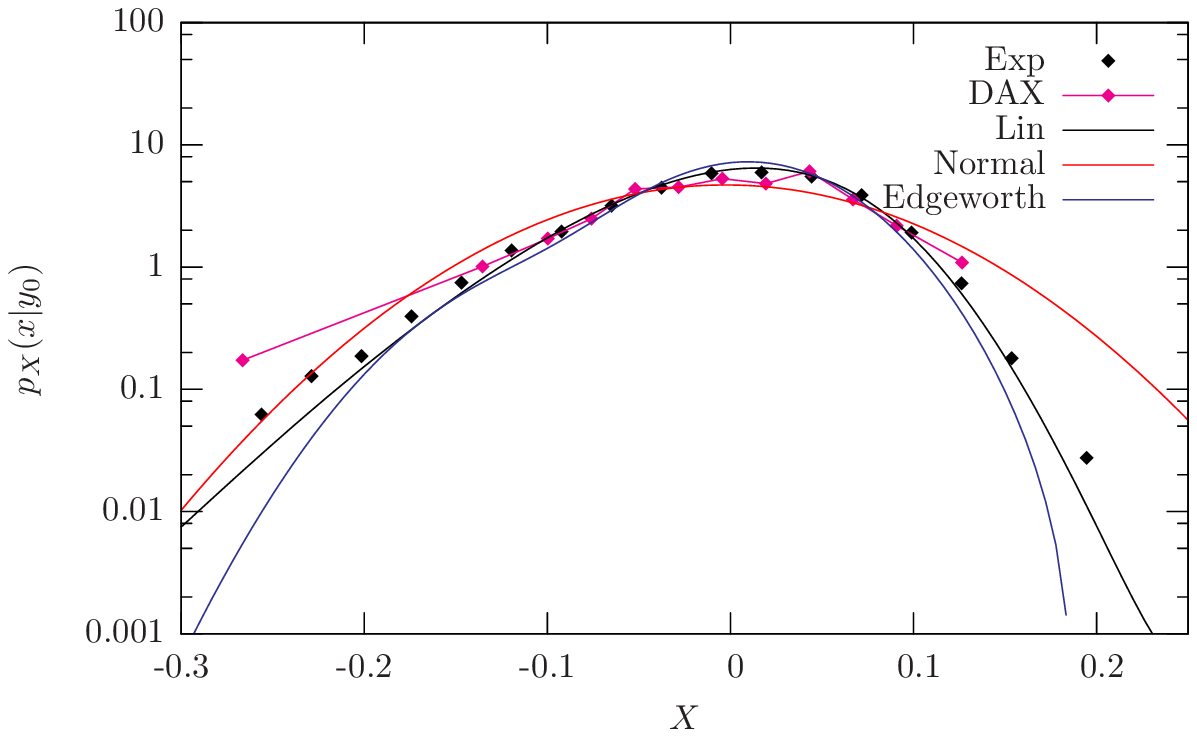}
    \end{center}
  \end{minipage}
  \begin{minipage}[b]{0.45\textwidth}
    \begin{center}
      \includegraphics[scale = 0.68]{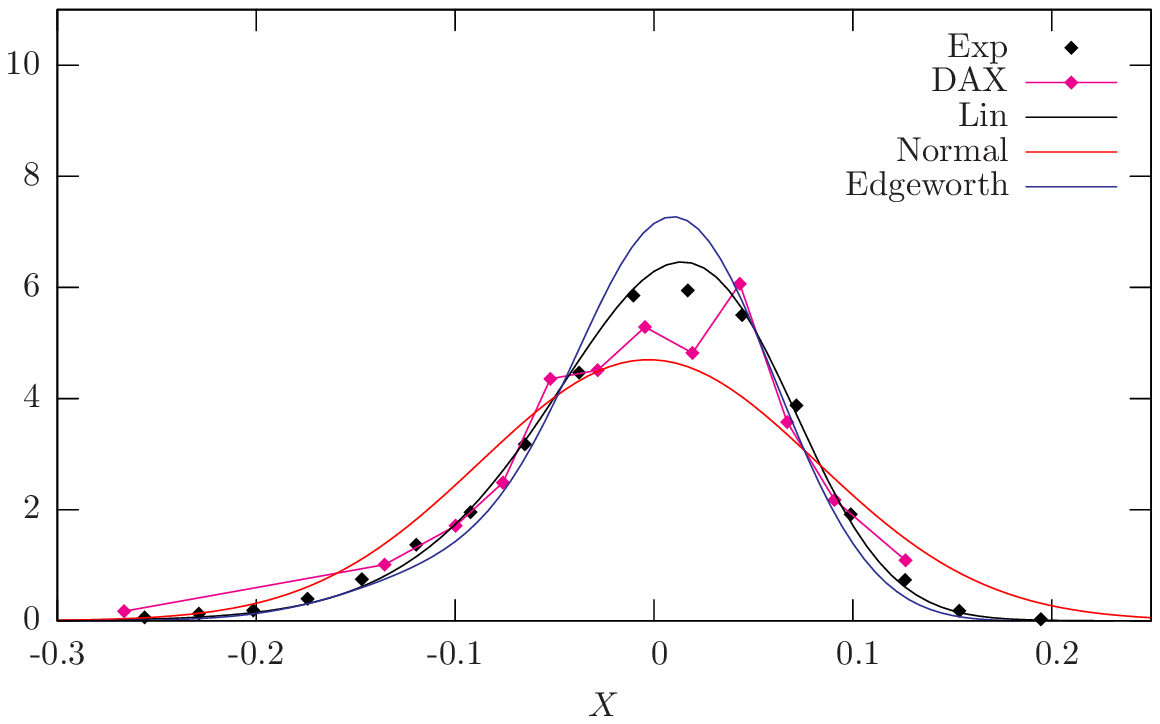}
    \end{center}
  \end{minipage}
\end{figure}
Finally, we present the comparison between the probability distributions for DAX30, see Fig.~\ref{fig:DAXScaling}, and 
for DJ Euro Stoxx 50, see Fig. ~\ref{fig:DJScaling}, computed according to the various models we have discussed in this
work. First row of both figures is obtained setting the time horizon equal to 25 trading days, 
while the second one corresponds to the 45 trading days horizon. 
The distributions are presented both in log-linear and linear scales, 
in order to allow the reader to appreciate the behavior on the tails and in the central region, respectively. 
The boxed-lines represent the empirical histograms (all the bins contain at least ten points), 
while black points correspond to the histograms from the MC simulation of the exponential Ornstein-Uhlenbeck model. 
Moreover, we report the Fourier transform of the characteristic function for the linear model (solid black line), 
the Normal Maximum Likelihood fit (solid red line) and the analytical solution \eref{eq:pXEdgeworth} 
(solid blue line). The Normal approximation is scarcely representative of the true empirical distribution, while
the exponential model captures in a quite effective way the leftward asymmetry, the fatter tails and the narrower 
central region. The Fourier transform of Eq.~\eref{eq:cflinmodel} and Eq.~\eref{eq:pXEdgeworth} are both candidates
to approximate the distribution of the exponential model. For the DAX30 index, the level of $\beta$ equal to $11.16\%$
allow us to be confident in a good performance of the linear model. This is indeed the case, as it is clearly shown
in Fig.~\ref{fig:DAXScaling}. The situation slightly worsen for the DJ Euro Stoxx 50 index, where $\beta=26.51\%$.
However, also in this case, the linear model performs better than the solution based on the Edgeworth 
expansion, that presents a narrower central peak and thinner tails. 
Moreover, for the r.h.s. of Eq.~\eref{eq:pXEdgeworth} positive definiteness is not guaranteed.
This is confirmed looking at the right tails in Fig.~\ref{fig:DAXScaling}, where the distributions 
in log-linear scale have to be truncated since the presence of negative values.\par        
Before concluding, it should be emphasized that further considerations 
about the capability of the linear model to reproduce known financial evidences
could be derived by comparing the theoretical predictions of the model for the fair prices of
options with the corresponding market option values. 
Actually, since our model has been mainly developed having in mind the time evolution of risk 
neutral distributions, this would allow to derive 
implied distribution functions over a given time horizon to be compared with the expectations
of the model. For example, in \cite{Bouchaud_Potters:2000,Potters_Cont:1998}, it is shown that the implied 
distribution functions corresponding to options with a time to maturity of one month have an empirical 
kurtosis value of the order 1, in good agreement with the model predictions for
$\beta$ values in the range $5\% \div 10\%$. This kind of analysis, which could be performed 
along the lines of \cite{Duffie:2000}, is presently
in progress and is left to a future publication.

%
%
\section{Conclusions and perspectives}\label{s:conclusions}  

In this paper we have dealt with the problem of the analytical characterization of the probability 
distribution of returns under an exponential Ornstein-Uhlenbeck dynamics for volatility. Such processes have been widely 
studied, in particular in the econophysics literature, from the point of view of multi-time scale properties,
leverage effect and stationary volatility distribution 
\cite{Bouchaud_Potters:2000,Dragulescu_Yakovenko:2002,Masoliver_Perello:2002,Mas_Per_An:2004,
Masoliver_Perello:2006,Cisana_etal:2007}. 
However, to the best of our knowledge, a systematic approach to the study of returns distribution is missing in 
literature. Our interest is referred to the possibility of a financial application in the context of option pricing
of a closed-form expression, even if approximated, for the probability distribution. The first attempt to obtain
such a result can be found in \cite{Masoliver_Perello:2006} under the hypothesis discussed in 
Section~\ref{s:model}, in particular under the assumption of a stationary regime for the hidden $Y$ process.  
We have generalized that result to the out-of-stationary regime, with the same constraint of the amplitude of volatility 
fluctuations, $k$, higher than its normal level, $m$ ($\lambda=k/m\gg 1$). 
Indeed, we have provided a closed-form expression in terms of the Edgeworth expansion.
This is one of the main results of our work. We have strongly tested our analytical result with MC simulations of the 
discrete Euler-Maruyama scheme of Eq.~\eref{eq:dY} and Eq.~\eref{eq:dX}. 
The goodness of the approximation has been checked with a careful analysis of the discretization 
step and MC number of paths for a quite reasonable choice of model parameters. 
We have found a good agreement for low values of $\beta$, the stationary variance of the log-volatility, while our
results worsen when $\beta$ increases even if $\lambda$ increases too. 
For this reason in Section~\ref{s:beta} we have explored the scenario opened by low $\beta$ values 
which allows us to expand the exponentials involved in the model up to first order in $Y$. 
Following the technique pioneered by Heston in \cite{Heston:1993}, we have solved exactly the Fokker-Planck 
backward equation associated with the linear model by means of a suitable trial solution. 
The full expression is reported in Section~\ref{s:beta} and is the main result of this work. 
Eq.~\eref{eq:mcC}-Eq.~\eref{eq:mcA} are quite involved and require some cares, as stressed in the text. 
Indeed they naturally arise in a complex domain and the presence of multi-valued square root and logarithmic
function introduces pain and angers of branching effects. 
They have been tackled following suggestions coming from the literature related to the Heston model. 
In particular, we have verified the smoothness of the real and imaginary part of the 
characteristic function for the parameters we consider. 
The final task of Fourier anti-transformation has been accomplished by means of FFT techniques. 
We have numerically tested the effectiveness of the analytical expression and we have found a perfect agreement. 
Moreover, the agreement between the results of the MC simulation of the discrete scheme of both
the complete and linear model make us confident of the numerical convergence of the exponential process for low 
$\beta$ values. 
The last section of this paper has been devoted to the comparison of analytical predictions with real market data.
The results of our analysis, in particular those for the German DAX30 and the Dow Jones Euro Stoxx 50 indexes, 
confirm the capability of the linear model to capture the statistical properties of the 
returns distribution for moderate values of $\beta$ and over appropriate time horizons.\par
As future development we plan to show how the SV model we have studied in this work can emerge 
in a quite natural way in the context of option pricing and risk management. 
In particular, we will discuss how to construct a hedged portfolio with associated 
underlying's dynamics following the exponential Ornstein-Uhlenbeck model. 
The knowledge of a closed-form expression for the characteristic function allows to implement 
a Carr~-~Madan-like approach and to calibrate the Heston-like option price 
formulae over the implied volatility surfaces. 
Moreover, we are interested in performing a numerical comparison between option prices 
obtained through full MC evaluation, our analytical formulae based on a linear model 
and other closed-form expressions available in literature
\cite{Perello_Sircar_Masoliver:2008,Filho_Rosenfeld:2004}.

%
%
\ack
We would like to thank D.Delpini and F.Piccinini for valuable discussion and helpful 
assistance in the later stage of this work.
We also acknowledge the anonymous referee for useful comments.
%
%
\appendix

%
%
\section{Derivation of $A(\omega_1,y_0,\tau)$, $B(\omega_1,y_0,\tau)$ and $C(\omega_1,y_0,\tau)$}\label{appA}

In order to derive approximate relations for the functions $A$, $B$ and $C$, we substitute the ansatz \eref{eq:ansatz}
in Eq.~\eref{eq:phiFPforward}. From relation \eref{eq:pXtransform} we argue that relevant information correspond
to small $\omega_2$ regime and moreover we look for a solution valid in large $\lambda$ regime. Expanding the exponentials
up to order $1/\lambda$ and equating the coefficients of $\omega_2$, $\omega_2^2$ and terms independent of $\omega_2$,
we find the following first order ODEs   
\begin{eqnarray}
  \label{eq:dotA}
  \dot{A} &= -2\theta(\omega_1)A + \frac{1}{2}\lambda^{2}\ ,\\
  \label{eq:dotB}
  \dot{B} &= -\theta(\omega_1)B+ \frac{2i\omega_{1}^{2}}{\lambda}A 
  -\frac{i\alpha\,\gamma\lambda}{k^2}+\lambda\rho\omega_{1}\ ,\\
  \label{eq:dotC}
  \dot{C} &= \frac{i\omega_{1}^{2}}{\lambda}B + \frac{1}{2}\omega_{1}^{2}+\frac{i\omega_1}{2\lambda}
  \ ,
\end{eqnarray}
with
\begin{equation}
  \theta(\omega_1)\doteq\frac{1}{2\beta}-i\rho\omega_1\ .
\end{equation}
The dot denotes a derivative {\it w.r.t.} $\tau$. The initial condition \eref{eq:initialphi} traduces in 
\begin{eqnarray}
  \label{eq:initialA}
  A(\omega_1,y_0,0)=&0\ ,\\
  \label{eq:initialB}
  B(\omega_1,y_0,0)=&-i\lambda y_0\ ,\\
  \label{eq:initialC}
  C(\omega_1,y_0,0)=&0\ .
\end{eqnarray}
Solutions of equations \eref{eq:dotA}, \eref{eq:dotB} and \eref{eq:dotC} with boundary conditions \eref{eq:initialA},
\eref{eq:initialB} and \eref{eq:initialC} read respectively
\begin{eqnarray}
  \label{eq:A}
  A(\omega_1,y_0,\tau)=&\frac{\lambda^2}{4\theta(\omega_1)}\left[1-\ue^{-2\theta(\omega_1)\tau} \right] \ ,\\
  \label{eq:B}
  B(\omega_1,y_0,\tau)=&\lambda\Bigg[-iy_0\ue^{-\theta(\omega_1)\tau}
    +\frac{i\omega_1^2}{2\theta^2(\omega_1)}\left(1-\ue^{-\theta(\omega_1)\tau}\right)^2 \nonumber \\
    &+\frac{1}{\theta(\omega_1)}\left(\rho\omega_1-\frac{i\alpha\gamma}{k^2}\right)
    \left(1-\ue^{-\theta(\omega_1)\tau}\right)\Bigg]\ ,\\
  \label{eq:C}
  C(\omega_1,y_0,\tau)=&\frac{i\omega_1}{2\lambda}\tau+\frac{\omega_1^2}{2}\tau\nonumber\\
  &+i\omega_1^2\Bigg[-iy_0\frac{1-\ue^{-\theta(\omega_1)\tau}}{\theta(\omega_1)}\nonumber\\
  &+\frac{i\omega_1^2}{2\theta^2(\omega_1)}\Bigg(\tau-2\frac{1-\ue^{-\theta(\omega_1)\tau}}{\theta(\omega_1)}
    +\frac{1-\ue^{-2\theta(\omega_1)\tau}}{2\theta(\omega_1)}\Bigg)\nonumber\\
    &+\frac{1}{\theta(\omega_1)}\Bigg(\rho\omega_1-i\frac{\gamma}{2\beta}\Bigg)\Bigg(\tau
    -\frac{1-\ue^{-\theta(\omega_1)\tau}}{\theta(\omega_1)}\Bigg)\Bigg]
  \ .
\end{eqnarray}
%

%
%

\end{document}